\documentclass[fleqn,usenatbib]{mnras}

\usepackage{newtxtext,newtxmath}
\usepackage[T1]{fontenc}

\DeclareRobustCommand{\VAN}[3]{#2}
\let\VANthebibliography\thebibliography
\def\thebibliography{\DeclareRobustCommand{\VAN}[3]{##3}\VANthebibliography}

\usepackage{graphicx}	% Including figure files
\usepackage{amsmath}	% Advanced maths commands
\def\sw{{\it Swift}~}
\usepackage[switch]{lineno}

\title[S5 1803+784 MWL flares]{Multi--wavelength flare observations of the blazar S5 1803+784}

\author[R. Nesci et al.]{
R. Nesci,$^{1}$\thanks{E-mail: roberto.nesci@inaf.it}
S. Cutini,$^{2}$
C. Stanghellini,$^{3}$
F. Martinelli,$^{4}$
A. Maselli,$^{5,6}$
V.M. Lipunov,$^{7}$
V. Kornilov,$^{7}$
\newauthor
R. R. Lopez,$^{8}$
A. Siviero,$^{9}$
M. Giroletti,$^{3}$
and M. Orienti$^{3}$
\\
$^{1}$INAF/IAPS, via Fosso del Cavaliere 100, 00133 Roma, Italy\\
$^{2}$INFN, Via Alessandro Pascoli, s.n.c, 06130, Perugia, Italy\\
$^{3}$INAF/IRA, Via P. Gobetti, 101, 40129, Bologna, Italy\\
$^{4}$Lajatico Astronomical Centre, Italy\\
$^{5}$INAF/OAR, Via Frascati 33, I-00078, Monte Porzio Catone (Roma), Italy\\
$^{6}$ASI/SSDC, Via del Politecnico snc, I-00133, Roma, Italy\\
$^{7}$Lomonosov Moscow State University, Physics Department, Sternberg Astronomical Institute, 119991 Moscow, Vorobievy hills, 1\\
$^{8}$Instituto de Astrofisica de Canarias Via Lactea, s/n E38205 - La Laguna (Tenerife), Spain\\
$^{9}$Department of Physics and Astronomy, University of Padova, Italy
}

\date{Accepted 2021-02-17. Received 2021-02-17; in original form 2020-11-12}

\pubyear{2021}

\begin{document}
\label{firstpage}
\pagerange{\pageref{firstpage}--\pageref{lastpage}}
\maketitle
%\linenumbers

\begin{abstract}

The radio, optical, and $\gamma$-ray light curves of the blazar S5 1803+784, from the beginning of the {\it Fermi} Large Area Telescope (LAT) mission in August 2008 until December 2018, are presented. The aim of this work is to look for correlations among different wavelengths useful for further theoretical studies.
We analyzed all the data collected by {\it Fermi} LAT  for this source, taking into account the presence of nearby sources, and we collected optical data from our own observations and public archive data to build the most complete optical and $\gamma$-ray light curve possible. 
Several $\gamma$-ray flares ($\mathrm{F>2.3~10^{-7} ph(E>0.1 GeV)~cm^{-2}~s^{-1}}$) with optical coverage were detected, all but one with corresponding optical enhancement; we also found two optical flares without a $\gamma$-ray counterpart. 
We obtained two {\it Swift} Target of Opportunity observations during the strong flare of 2015. Radio observations performed with VLBA and EVN through our proposals in the years 2016-2020 were analyzed to search for morphological changes after the major flares.
The optical/$\gamma$-ray flux ratio at the flare peak varied for each flare. Very minor optical V-I color changes were detected during the flares. The X-ray spectrum was well fitted by a power law with photon spectral index $\alpha$=1.5, nearly independent of the flux level: no clear correlation with the optical or the $\gamma$-ray emission was found. The $\gamma$-ray spectral shape was well fitted by a power law with average photon index $\alpha$= 2.2.
These findings support an Inverse Compton origin for the high-energy emission of the source, nearly co-spatial with the optically emitting region. The radio maps showed two new components originating from the core and moving outwards, with ejection epochs compatible with the dates of the two largest $\gamma$-ray flares.

\end{abstract}


%%%%%%%%%%%%%%%%%%%%%%%%%%%%%%%%%%%%%%%%%%%%%%%%%%

%%%%%%%%%%%%%%%%% BODY OF PAPER %%%%%%%%%%%%%%%%%%

\section{Introduction} 
 \label{Intro} %section 1

BL Lacertae objects are a subclass of Active Galactic Nuclei (AGN) characterized by fast and large flux variations. Their emission ranges from radio frequencies up to $\gamma$-rays, and their spectral energy distribution (SED) in a Log($\nu$) {\it vs} Log($\nu \cdot F_{\nu}$) plane may be described by a double bell shape: the first peak is located between the far infrared and X-ray frequencies, while the second one is always at much higher energies (from X-ray to $\gamma$-ray energies).
A current interpretation of this double bell shape is that synchrotron radiation from relativistic electrons in a highly collimated jet is responsible for the low-frequency peak, while inverse Compton radiation from these electrons impacting local photons produces the higher-frequency peak.
This model is likely an oversimplification with respect to reality: the black hole at the centre of the host galaxy is surrounded by an accretion disk, a hot corona, and the relativistic jet approximately directed along the black hole rotation axis is likely structured at least in an inner part (spine) and an outer envelope (sheath) with different physical conditions \citep{2000A&A...358..104C}.
Such a differentiated structure implies that the radiation at  different frequencies may come from physically different spatial regions: flux variations at different frequencies may therefore reach the observer at different times. Thus multi-wavelength simultaneous observations are an efficient tool to explore the physical structure of the emitting regions. 

This paper is an observational contribution to the study of S5 1803+784, a radio-selected BL Lac object \citep{1981ApJ...247L..53B} at z=0.683 \citep{1996ApJS..107..541L}; a detailed modeling of the emission processes and morphology is beyond the scope of this paper.
S5 1803+784 is characterized by large variations in the optical range \citep{2002AJ....124...53N, 2012AcPol..52a..39N} and is well detected at $\gamma$-ray energies by the {\it Fermi} Large Area Telescope (LAT) \citep{2009ApJ...697.1071A}, from the first LAT Catalog of AGN \citep{2009ApJ...700..597A}, up to the most recent one \citep{2020ApJ...892..105A}. 
 In the taxonomical classification scheme of BL Lac objects by  \citet{1995ApJ...444..567P} the source is  Low Synchrotron Peaked (LSP), with the synchrotron emission peaked around 10$^{13}$ Hz: this is clearly shown by its broadband SED \citep{2002AJ....124...53N}. 

In this paper we present the light curves of the source in the radio (15 GHz), optical (R$_C$) and $\gamma$-ray bands from year 2008 to 2018, with snapshots in the X-ray and optical bands by the {\it Neil Gehrels Swift Observatory} X-ray Telescope (XRT) and the UltraViolet and Optical Telescope (UVOT), to search for correlations among different wavelengths. Furthermore we present Very Long Baseline Interferometry (VLBI) radio observations performed after two large flares, to look for morphological changes in its inner jet structure. 

\section{Observational data}
\subsection{Optical Photometry}
\label{sec:maths} % used for referring to this section from elsewhere
We have been monitoring this source in the optical band since 1996 \citep{2002AJ....124...53N,2012AcPol..52a..39N} despite with some large gaps. It was serendipitously observed by the MASTER robotic network \citep{2010AdAst2010E..30L} since 2010. Due to its very northern position in the sky it was not covered by the Catalina Sky Survey \citep{2009ApJ...696..870D}; since July 2011 this source became a target of the blazar monitoring program by the KAIT telescope \citep{2003PASP..115..844L}.

In this paper we used our observations taken with several telescopes since 2009: the 50cm F/4.5 at Vallinfreda, the 30cm F/4.5 at Greve in Chianti, the 1.5m telescope at the Loiano Observatory, the 23cm F/10 of the Department of Physics of La Sapienza University, the 50cm F/4.4 at the Lajatico Astronomical Centre. All these telescopes are equipped with CCD cameras and Cousins R filters. Magnitudes were derived using aperture photometry with IRAF/apphot\footnote{\small IRAF - Image Reduction and Analysis Facility, distributed by NOAO, operated by AURA, Inc. under agreement with the US NSF.}, using always the same comparison sequence \citep{2002AJ....124...53N}. 

\begin{figure*}
\centering
\includegraphics[height=13cm]{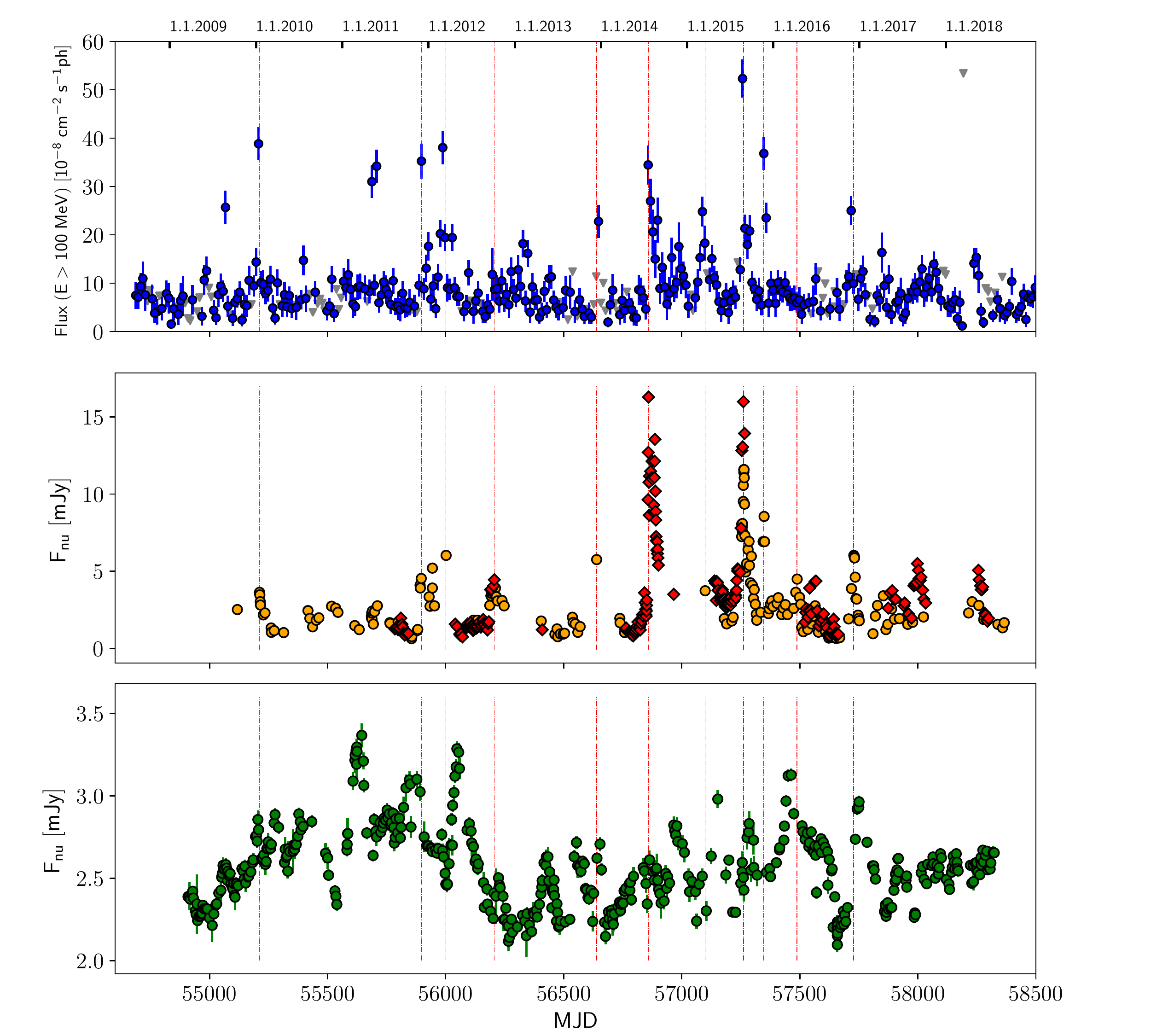}
\caption{The combined light curve of S5 1803+784 in linear scale: lower scale in MJD, upper scale in calendar years. Upper panel, $\gamma$-ray, binned at 10-day intervals: blue points are detections, grey triangles are upper limits; middle panel, optical: orange points are our own data, red points are KAIT public data; lower panel, radio 15 GHz from OVRO monitoring. Vertical lines mark the flares discussed in the text.}
\label{curva2}
\end{figure*}

In past years \citep{2002AJ....124...53N,2012AcPol..52a..39N}, the color indexes (V-R and R-I) of the source showed only small variations, with a mild bluer-when-brighter behaviour. This is also shown by the present data set (see Table \ref{opt_col},  from which we derive an average increase of 0.04 mag in R-I for a 1 mag increase in R), and makes us confident of the reliability of converting V magnitudes into R ones with a fixed color index for the purpose of building a denser historical light curve. Therefore we included in our data set the V magnitudes from the 30cm F/10 telescope of the Foligno Observatory and from the 40cm F/8 telescope of the Royal Observatory of Belgium \citep{2015ATel.7988....1L}, converting their V magnitudes into R$_C$, adopting V-R=0.50 mag. For the KAIT observations, which are unfiltered and use the USNO-B1 \citep{2003AJ....125..984M} R2 magnitudes as reference, we applied an average correction of -0.20 mag to their psf magnitudes, based on the average offset of our photometric sequence stars with respect to the USNO-B1 catalog: this value is within the quoted systematic uncertainty given in the KAIT database.
For the MASTER observations, which are also unfiltered and expressed as a nearly R$_C$ magnitude, we verified a good consistency with our magnitudes for the few nights of simultaneous observations. The resulting overall R$_C$ light curve is shown in the middle panel of Fig.~\ref{curva2}.

\begin{table}
\caption{Optical colors from la Lajatico (LA) and Loiano (LO) Observatories multifilter observations. Date in Column 1, R$_C$ in column 2, V-R$_C$ in column 3, R$_C$-I$_C$ in column 4, V-I$_C$ in column 5, Observatory in column 6. }             % title of Table
\label{opt_col}      % is used to refer this table in the text
\centering                          % used for centering table
\begin{tabular}{c c c c c c c}        % centered columns (6 columns)
\hline\hline                 % inserts double horizontal lines
date& date  &   R & V-R & R-I & V-I  &  tel \\
    &MJD   & mag& mag  & mag & mag  &       \\
\hline                        % inserts single horizontal line
2011-07-21 &55763.84&15.85 &0.49 &0.70 & 1.19 & LO\\
2011-07-22 &55764.83&15.88 &0.50 &0.70 &1.20  &LO\\
2015-08-21 &57255.85&14.19 &0.49 &0.59& 1.08&  LA\\
2015-08-26 &57260.82&13.91 &0.50 &0.59 &1.09 & LA\\
2015-08-28 &57262.84&13.76 &0.51 &0.64 &1.15  &LA\\
2015-08-31 &57265.86&13.97 &0.46 &0.61 &1.07  &LA\\
2015-09-05 &57270.86&14.66 &0.48 &0.68 &1.16  &LA\\
2015-09-09 &57274.84&14.55 &0.48 &0.68 &1.16  &LA\\
2015-09-19 &57284.87&14.30 &0.53 &0.65 &1.18  &LA\\
2015-09-21 &57286.81&14.58 &0.48 &0.69 &1.17  &LA\\
2015-10-11 &57306.81&15.02 &0.45& 0.68& 1.13&  LA\\
2015-10-22 &57317.75&15.75 &0.51& 0.71& 1.22 & LA\\
2015-11-24 &57350.74&14.23 &0.50& 0.62& 1.12  &LA\\
2016-05-21 &57729.92&15.97 &0.49& 0.69& 1.18  &LA\\
2016-06-26 &57565.07&15.29 &0.49& 0.78& 1.27  &LA\\
2016-07-01 &57570.07&15.48 &0.61& 0.63& 1.24  &LA\\
2016-07-09 &57578.92&16.33 &0.35& 0.66& 1.01  &LA\\
2016-12-15 &57737.80&15.13 &0.49& 0.00& 0.00 &LA\\
2016-12-29 &57751.78&15.77 &0.46&0.00 &0.00  &LA\\
\hline                                   %inserts single line
%\text{Notes - LO=Loiano; LA=Lajatico}
\end{tabular}
\end{table}

This light curve shows a baseline flux level between 1 and 2 mJy, with several flares, and some short time intervals with flux below 1 mJy.

The definition of flare is somewhat arbitrary, since it depends on the assumed quiescent level of the source. From the optical light curve we derived an average quiescent level of 1.5 mJy (R=16.0 mag), and we considered a flare as a rise and fall episode in the light curve with a peak value higher than 3 mJy (R=15.2 mag). 
For the better sampled flares we measured the duration and the slopes of the rising and falling branches. The duration was defined as the time interval during which the source was above the average level (R$_C$=16.0 mag): the uncertainties of these intervals depend on the sampling of the light curve. The slopes (mag/day) were derived with a linear fit to the manually selected relevant data.
A total of 11 flares were identified in this way and are listed in Table \ref{opt_flares}. The simultaneous presence of a $\gamma$-ray flare (see Section 2.3 below) is given in the last column.

\begin{table}
\caption{Optical flares data: date of the maximum (column 1), the peak R$_C$ magnitude (column 2), the flare duration (column 3), the slope of the rising (column 4), and of the dimming  branch (column 5).}             % title of Table
\label{opt_flares}      % is used to refer this table in the text
\centering                          % used for centering table
\begin{tabular}{c c c c c c}        % centered columns (4 columns)
\hline\hline                 % inserts double horizontal lines
date &peak R$_C$& duration & rising     & falling & $\gamma$ flare\\
MJD & mag & days       &mag/d   &mag/d &      \\
\hline                        % inserts single horizontal line
55211  & 15.0 &  45 &       & 0.07  & yes \\
55897  & 14.7 &     &       &       & yes \\ 
56001  & 14.4 &     &       &       & yes \\
56206  & 14.5 & 60  & 0.076 &       & no  \\
56639  & 14.5 &     &       &       & yes \\
56859  & 13.9 &  45 & 0.22  &       & yes \\
57099  & 15.0 &     &       &       & yes \\
57262  & 13.7 &     & 0.11  &       & yes \\
57348  & 14.0 &     &       &       & yes \\
57488  & 14.8 &     &       &       &  no \\
57728  & 14.4 &  30 & 0.08  &  0.10 & yes \\
\hline                        % inserts single horizontal line
\end{tabular}
\end{table}

\subsection{Optical spectroscopy} %section 2.2

Only three spectra of this source have been published so far: one taken in 1987 \citep{1996ApJS..107..541L} showing the MgII 2790\AA ~emission line with equivalent width (E.W.)=8.8\AA; one taken in 1996 \citep{2001AJ....122..565R} that reports for the MgII line an E.W=2.8 \AA; the last taken in 2018 \citep{2020MNRAS.497...94P} with E.W.=8.3 \AA, very similar to the value of 1987. The magnitude of the source in 1987 is unknown, while from our monitoring we know that it was  faint in 1996 (R=16.2 mag), as well as in 2018 (R=15.9 mag).

\begin{figure}
\centering
\includegraphics[height=8cm,angle=0]{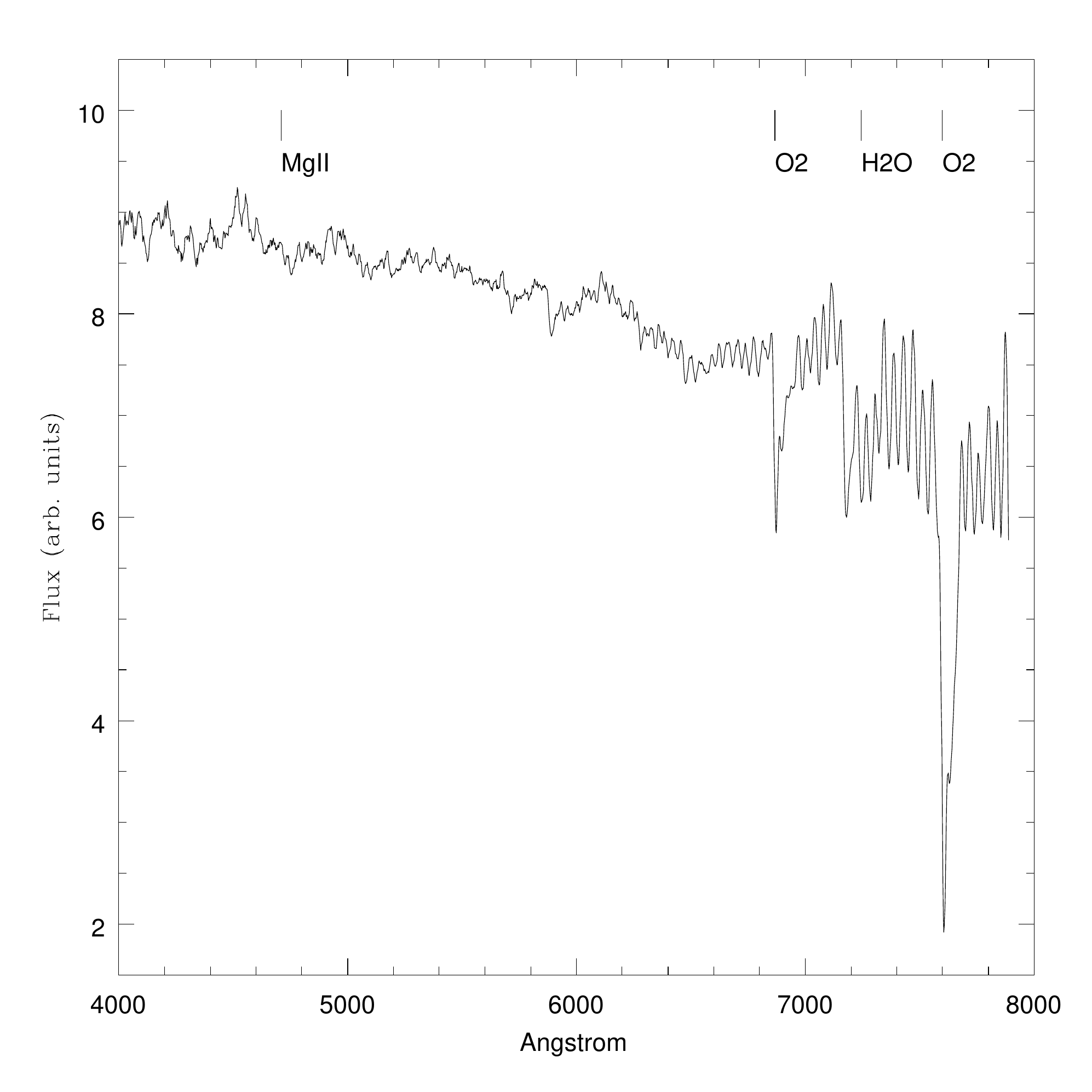}
\caption{The spectrum of S5 1803+784 from the Asiago Observatory, corrected for instrumental response and smoothed with a running mean of 7 pixels, in arbitrary flux units. Telluric Oxygen (O$_2$) and water absorption bands (H$_2$O) are well evident; the expected position of the MgII emission is also marked. The apparent emission around 4525 \AA ~is a residual from night-sky emission lines.
}
\label{spett}
\end{figure}

We obtained a spectrum of the source on 27 August 2015 (MJD 57261), when it was near the maximum level (R=13.8 mag), with the 1.22m telescope of the Asiago Observatory equipped with an Andor iDus DU440 camera and a Boller \& Chivens spectrograph with a 300 gr/mm grating: the dispersion was 2.3 \AA/pixel and the spectral resolution 9.5 \AA, derived from the FWHM of the night-sky emission lines. Wavelength calibration was made with a He-Fe-Ar lamp and instrumental response calibration with the star HR 7596. The blazar spectrum in arbitrary flux units is shown in Fig.\ref{spett}. 

The S/N ratio was about 20, allowing the detection of an emission line of E.W= 3 \AA.  No such line was present at 4710 \AA, the expected position of the MgII line, and the spectrum showed no features besides the telluric ones. The spectral slope was fully consistent with that derived from our VRI photometry. 

\subsection{ {\it Fermi}-LAT data analysis} %section 2.3

The data analysis was performed following the {\it Fermi}-LAT collaboration recommendations for point-source analysis\footnote{\small{https://fermi.gsfc.nasa.gov/ssc/data/analysis/documentation/ \\
Pass8\_usage.html}} 
and is briefly described as follows. We built the historical light curve collecting almost 10 years of data available from the {\it Fermi}-LAT archive, starting from 4 August 2008 up to 31 December 2018. We analyzed the {\it Fermi}-LAT data using the FermiTools package version 1.2.1 available from the Fermi Science Support Center (FSSC)\footnote{\small{https://fermi.gsfc.nasa.gov/ssc/}} through the Conda package manager hosted on GitHub\footnote{https://github.com/fermi-lat/Fermitools-conda/} and the P8R3$\_$SOURCE$\_$V2 instrument response functions \citep{pass8}. We selected a Region of Interest (ROI) with a radius of 15 degrees  centered at the position of S5 1803+784 and we used only events belonging to the “source” class. The binned maximum likelihood was performed using as background model the 4FGL sources lying in a  ROI with a radius of 20 degrees (4FGL; \citealt{2020ApJS..247...33A}) and the isotropic and Galactic diffuse emission components (iso$\_$P8R3$\_$SOURCE$\_$V2$\_$v1.txt and gll$\_$iem$\_$v07.fits\footnote{\small{https://fermi.gsfc.nasa.gov/ssc/data/access/lat/\\
BackgroundModels.html}}, respectively). We fixed the spectral indexes of background sources to the 4FGL values, but the fluxes  of sources within a radius of 10 degrees from the source of interest were left free to vary. Light curves were derived by dividing the data in bins of 10-day duration for the historical and 1-day and 3-day bins for the flaring epochs. 
We defined as flaring state a flux above 100 MeV larger than $\mathrm{2.3\times10^{-7} ph~cm^{-2}s^{-1}}$ in a 10-day bin, three times the average value of fluxes in the 10-day light curve. Each bin of the light curves and of the spectra was obtained using a power-law model for the source of interest.
For each bin we extracted the Test Statistic (TS; \citealt{1996ApJ...461..396M}) defined as:
\begin{math}
TS=-2(lnL_0-lnL_1)
\end{math}
where $L_0$ is the value of the likelihood calculated in the null hypothesis (no point source in ROI) and $L_1$  is the value of the likelihood when the source of interest is included (after parameter optimization).  We calculated the upper limit (UL) at $2\sigma$ confidence level using the Bayesian computation  \citep{1983NIMPR.212..319H} and integrated UL when the TS$<$1 and TS$<4$ respectively,
the number of predicted photons Npred$<$3, or the percentage error on the flux greater than 50\%.
 The light curve in the 0.1-300 GeV range, binned at 10-day intervals, is shown in the upper panel of Fig. \ref{curva2}: the light curve shows a flat, oscillating behaviour with a typical flux level of $\mathrm{0.75\times10^{-7} ph~cm^{-2} s^{-1}}$ and several large flares, irregularly spaced, listed in Table \ref{Gamma10}. The flux in the largest flares was nearly 10 times higher than the quiescent state. Thanks to the regular monitoring of the satellite, for all the $\gamma$-ray flares it was possible to measure a duration, which ranges between 20 and 90 days, with an uncertainty of 3 days. In Table \ref{Gamma10} we report the $\gamma$-ray flare details.

\begin{table*}
\caption{$\gamma$-ray flare data: the MJD of the peak emission (column 1), the peak flux value (column 2),  the flare duration (column 3), the corresponding optical flux (column 4), the optical/ $\gamma$-ray flux ratio (column 5, $\times10^{7}$)}.             % title of Table
\label{Gamma10}      % is used to refer this table in the text
\centering                          % used for centering table
\begin{tabular}{c c c c c}        % centered columns (4 columns)
\hline\hline                 % inserts double horizontal lines
date &     peak-flux        & duration & R$_C$ & Opt/gamma-ray \\
MJD  & ph~cm$^{-2}$~s$^{-1}$&   days   & mJy   &             \\
\hline                        % inserts single horizontal line
55067 & 2.6x10$^{-7}$ & 40 & $-$   &   $-$  \\
55207 & 4.0x10$^{-7}$ & 60 & 3.6 & 0.90 \\
55707 & 3.5x10$^{-7}$ & 80 & 2.7 & 0.77 \\
55897 & 3.8x10$^{-7}$ & 90 & 4.5 & 1.18 \\
55987 & 3.6x10$^{-7}$ & 40 & 6.0 & 1.66 \\
56647 & 2.3x10$^{-7}$ & 20 & 5.7 & 2.48 \\
56857 & 3.6x10$^{-7}$ & 60 &12.3 & 3.41 \\
57257 & 5.0x10$^{-7}$ & 40 &11.6 & 2.32 \\
57347 & 3.7x10$^{-7}$ & 30 & 8.5 & 2.30 \\
57717 & 2.5x10$^{-7}$ & 20 & 6.0 & 2.40 \\
\hline                        % inserts single horizontal line
\end{tabular}
\end{table*}

\begin{table*}
\caption{data from \citealt{swift} observations: the date (column 1), the MJD (column 2), the last digits of the observation identifier (column 3),  the exposure time (column 4),  the total counts (column 5), the  net count rate (column 6), the photon index (column 7),  the flux in the 0.3-10 kev range assuming a power-law spectrum and the Galactic absorption (column 8).}             % title of Table
\label{xray-flares}      % is used to refer this table in the text
\centering                          % used for centering table
\begin{tabular}{c c c c c c c c c }        % centered columns (4 columns)
\hline\hline                 % inserts double horizontal lines
date      &date    &obs&exp&counts& count rate   &Photon Index &Flux 10$^{-12}$    & R$_C$\\
          &MJD     &id & s &      & cts/s        &             & erg cm$^{-2}$ s$^{-1}$ & mag \\
\hline
2009-06-07&54989.93& 04&9337&710&0.075$\pm$0.003&1.43$\pm$0.06&4.08$\pm$0.25&16.29\\
2009-10-13&55117.00& 05&5097&269&0.052$\pm$0.003&1.47$\pm$0.10&3.05$\pm$0.30&15.40\\
2011-02-23&55615.11& 08&1274&100& 0.078$\pm$0.008&1.53$\pm$0.17&4.36$\pm$0.78&15.98\\
2011-02-23&55615.18& 09&2547&162&0.063$\pm$0.005&1.38$\pm$0.14&3.81$\pm$0.54&15.98\\
2011-05-05&55686.49& 11&3696&141&0.038$\pm$0.003&1.53$\pm$0.15&2.81$\pm$0.45&15.67\\
2011-05-06&55687.62& 12&3654&202&0.055$\pm$0.004&1.48$\pm$0.12&3.00$\pm$0.38&15.50\\
2011-05-07&55688.63& 13&2392&142&0.059$\pm$0.005&1.65$\pm$0.13&2.68$\pm$0.36&15.50\\
2011-05-08&55689.29& 14&4335&261&0.059$\pm$0.004&1.68$\pm$0.10&2.58$\pm$0.24&15.50\\
2011-05-09&55690.63& 15&2270& 81&0.035$\pm$0.004&1.74$\pm$0.24&2.36$\pm$0.68&15.44\\
2012-01-04&55930.31& 16&1394& 61&0.043$\pm$0.006&1.70$\pm$0.30&2.49$\pm$0.95&15.09\\
2012-01-08&55934.05& 17&4350&223&0.051$\pm$0.003&1.30$\pm$0.11&2.90$\pm$0.31&15.31\\
2015-08-23&57257.11& 18&1743&120&0.068$\pm$0.006&1.50$\pm$0.15&3.43$\pm$0.53&14.13\\
2015-09-09&57274.05& 19&3911&264&0.067$\pm$0.004&1.50$\pm$0.09&3.29$\pm$0.30&14.55\\
\hline
\end{tabular}
\end{table*}

\subsection{X-ray and Ultraviolet photometry} 
\label{IP} %section 2.4

We used X-ray data from the X-Ray Telescope (XRT) on board the {\it Neil Gehrels Swift Observatory} (\citealt{swift}), retrieved from the \sw archive and acquired from 2009 to 2015; these include data corresponding to the August 2015 flare, respectively 5 days before  and 11 days after the the optical maximum.
All data were processed with the XRTDAS software package (v.3.6.0), developed at the Space Science Data Center (SSDC) of the Italian Space Agency (ASI) and distributed within the HEASoft package (v.6.28) by the NASA High Energy Astrophysics Archive Research Center (HEASARC). All the XRT observations were carried out in the most sensitive photon counting (PC) readout mode. Event files were calibrated and cleaned applying standard filtering criteria with the {\sc xrtpipeline} task and using the latest calibration files available in the \sw CALDB distributed by HEASARC. Events in the energy range 0.3--10 keV with grades 0--12 were used in the analysis, and the exposure maps were also created with {\sc xrtpipeline}. Source detection was carried out using the detection algorithm {\sc detect} within {\sc ximage}, also providing the count rate corrected for psf, vignetting, and exposure. 

For each observation, source events were extracted from a circle with a radius of 20 pixels centered at the source coordinates; a concentric annulus with radii of 50 and 80 pixels was used for the background. These region files were then used as input to the {\sc xrtproducts} task to obtain high-level scientific products: among these, the spectrum and ancillary files were used to perform a spectral analysis with {\sc xspec} (v.12.11.1). We fitted these spectra using a power-law model, fixing the absorption at the Galactic value (n$_H$= 3.4~10$^{20}$ cm$^{-2}$), and we computed the flux in the 0.3--10 keV band with its 1~$\sigma$ error.

To compare the X-ray and optical luminosity we looked at the Swift-UVOT simultaneous observations. Aperture photometry of the UVOT images was made with the UVOT on-line analysis tool available from the SSDC (uvotimgqlprocver v1.14). The source radius was 5 arcsec and the sky level was evaluated within a concentric corona between 27 and 35 arcsec distant from the source. Unfortunately only in 4 pointings were all 6 UVOT filters used, but we found that the optical spectral slope was substantially constant (-0.57 $\pm$0.03 in the log($\nu$) vs log($\nu$ F($\nu$) plane), and therefore we could compute by extrapolation the corresponding R$_C$ magnitudes for all the UVOT pointings. In some cases we also had nearly simultaneous R$_C$ observations from the ground, confirming the reliability of our extrapolation procedure within 0.2 mag. 

Our results from the analysis of the \citealt{swift} data are reported in Table \ref{xray-flares}. The last column is the corresponding R$_C$ magnitude. Observations with relatively poor exposure are also reported.
Overall the XRT fluxes changed by a factor $\sim$2, and the spectral slope had an average value of 1.51 $\pm$ 0.13, with only marginal evidence (correlation coefficient r=-0.54) of hardening at higher fluxes. 

\section{Radio observations} %section 3
To explore if the strong flares of the years 2014 and 2015 produced detectable changes in the structure of the inner part of the source, we obtained radio images of the core of S5 1803+784 at sub-milliarcsecond  resolution.
Our observations were performed with the Very Long Baseline Array (VLBA) and the European VLBI Network (EVN) at 15, 22, and 43 GHz in six epochs, spanning from June 2016 to March 2018,  as listed in Table \ref{vlbi1}. 
We also collected radio data at 15 GHz from the 40m Telescope of the Owens Valley Radio Observatory (OVRO; \citealt{ovro}) public archive  to explore the possibility of activity simultaneous to the optical and $\gamma$-ray flares. The radio data are shown in the bottom panel of Fig. \ref{curva2}. We did not find any obvious evidence of correlations.

\subsection{Data reduction} %section 3.1
Amplitude and phase calibration of the VLBA data were carried out with the AIPS data reduction package in a standard way following the guidelines given in appendix C of the AIPS cookbook:  the typical flux error was 5\%. Data from the third VLBA epoch were of lower quality especially at 43 GHz, maybe for bad weather conditions at some antennas. 
A-priori calibration of antenna gains proved to be infeasible for EVN data, and we could not obtain reliable flux density values. We relied on several iterations of imaging and self-calibration (amplitudes and phases) to obtain the relative position of the components.
Model fitting of the visibilities was performed with the task MODELFIT available in the Caltech software package DIFMAP. 

The morphological radio structure of S5 1803+784 at milliarcsecond (mas) scale can be described as a diffuse emission from an opening jet, directed East-West, with knots and brightness enhancements in several locations along the jet (see e.g. \citealt{2008A&A...483..125R} and references therein). Evidence in support of a helical jet has been presented by \citet{2010A&A...511A..57B} and \citet{2018MNRAS.478..359K} from a long term monitoring at different frequencies.

This complex structure makes it difficult to compare observations at different epochs observed with different baseline distributions, thus different angular resolution and different sensitivity to diffuse emission. In particular, our 2016 observations were made with the full VLBA array (10 antennas) while in the 2017 sessions two antennas were missing (Los Alamos and Pie Town in March, St. Croix and Pie Town in November). To have consistency in the components found with MODELFIT, it is important to minimize the effect of the different baseline distributions, especially at the lowest frequency where the detection of the diffuse emission is more relevant. Therefore at 15 GHz, solely for the MODELFIT procedure, the UV planes of the data were matched as much as possible, excluding 2 antennas from the 2016 data sets and cutting as needed the inner baselines. %The difference from the original UV coverage and that used for the modefit is shown in Fig.4.

VLBA additional data at 15 GHz (7 epochs) from the MOJAVE project (\citealt{2018ApJS..234...12L}) on 2019 and 2020 were added to the analysis. The typical UV coverage of the MOJAVE data was significantly different from ours, and generally better. Calibrated visibilities were downloaded from the MOJAVE archive, imaged and model fitted again in AIPS and DIFMAP. 

\subsection{Morphological analysis} %section 3.2

\begin{figure}
\centering
\includegraphics[height=6cm,angle=0]{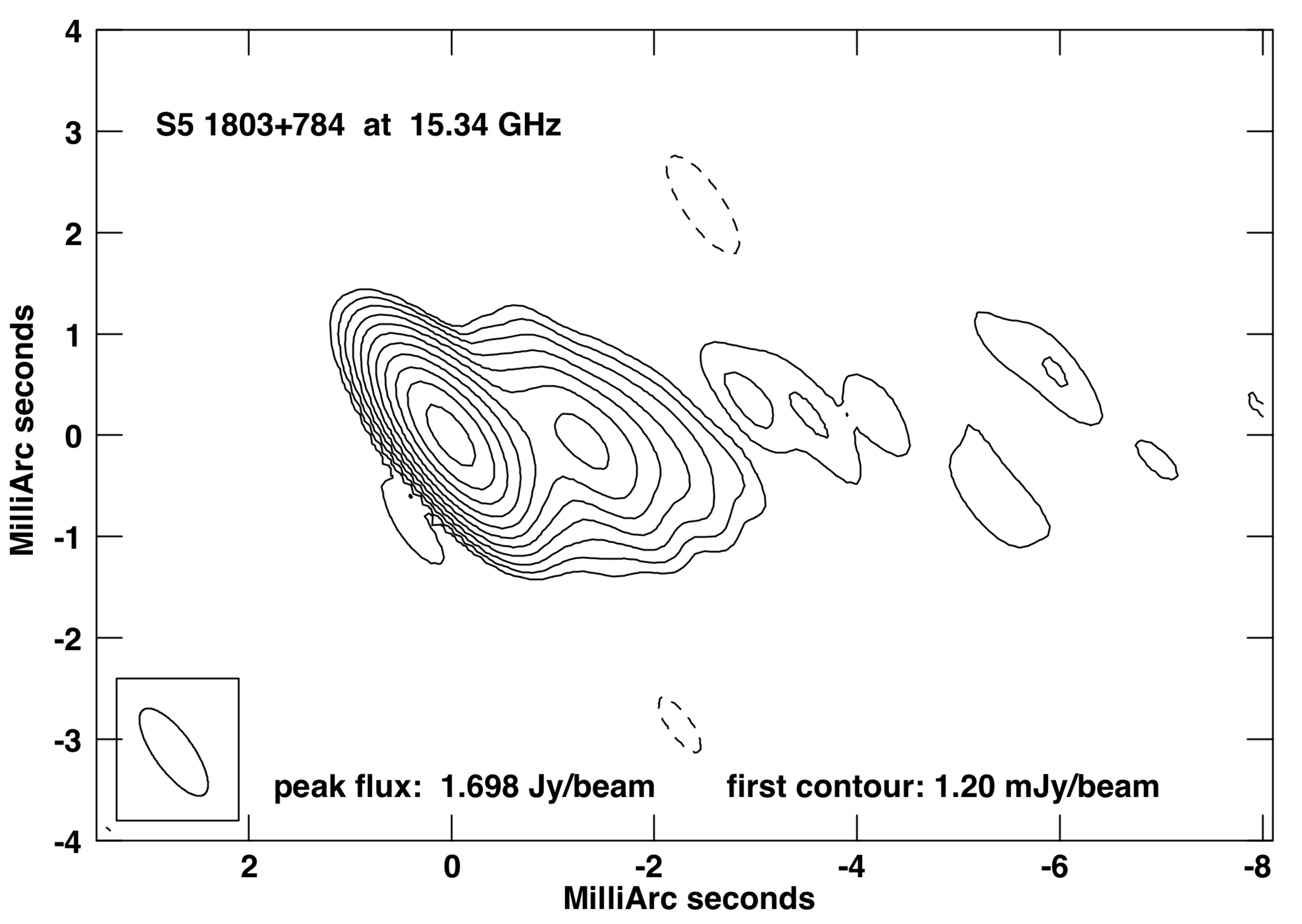}
\caption{The radio map at 15 GHz on 4 September 2016 at 1.03 x 0.38 mas$^2$ resolution from VLBA. Contours increase by a factor of 2.}
\label{r2}
\end{figure}

\begin{figure}
\centering
\includegraphics[height=8.8cm,angle=-90]{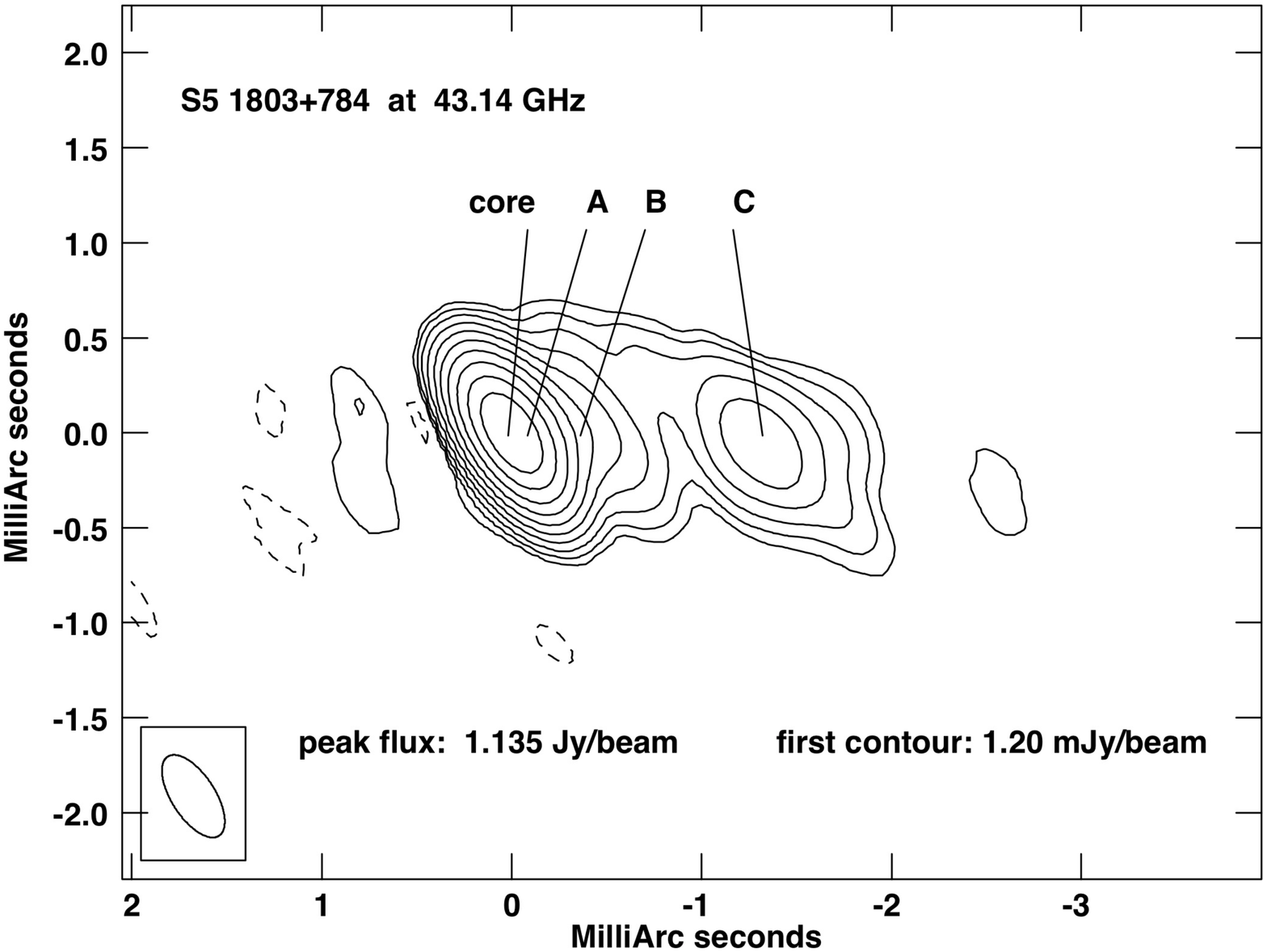}
\caption{The radio map at 43 GHz on 4 September 2016 at 0.50 X 0.23 mas$^2$ resolution from VLBA. Contours increase by a factor of 2.}
\label{r3}
\end{figure}

\begin{figure}
\centering
\includegraphics[height=5cm,angle=0]{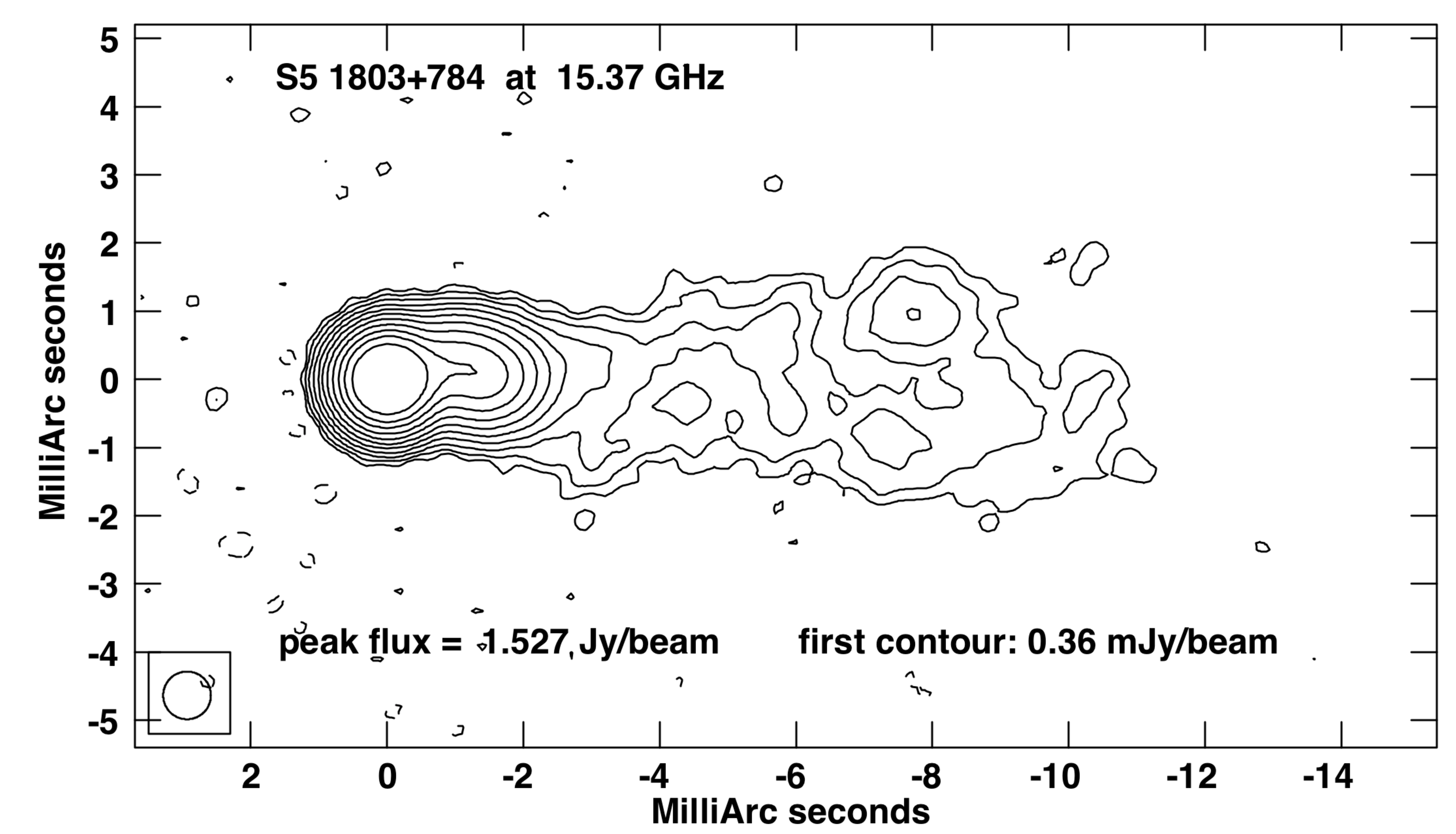}
\caption{The radio map at 15 GHz at 0.77x0.65 mas$^2$ resolution recomputed by us using MOJAVE data of the years 2019-2020. Contours increase by a factor of 2.}
\label{r4}
\end{figure}

In Fig. \ref{r2} we show our VLBA map at 15 GHz taken on 4 September 2016, while in Fig.~\ref{r3} we show our map at 43 GHz, which has a higher resolution. Fig.~\ref{r4} was built by us reprocessing all the public MOJAVE data at 15 GHz of the years 2019-2020, stacking the seven images together, and shows the source morphology at a larger scale.

Given the morphology of this radio source, the attempt to describe the parsec scale emission with a limited number of discrete components gives a rough approximation of the real structure. Furthermore, the choice of the number of components to describe the source is not trivial, especially when trying to compare the structure at various epochs. Additional caution is needed to compare the morphology at different frequencies, as different angular resolution, different sensitivity to diffuse emission, and different intrinsic opacity may introduce shifts in the positions and delays in the estimated proper motion.

\begin{figure} %figure 6
\centering
\includegraphics[height=8cm,angle=0]{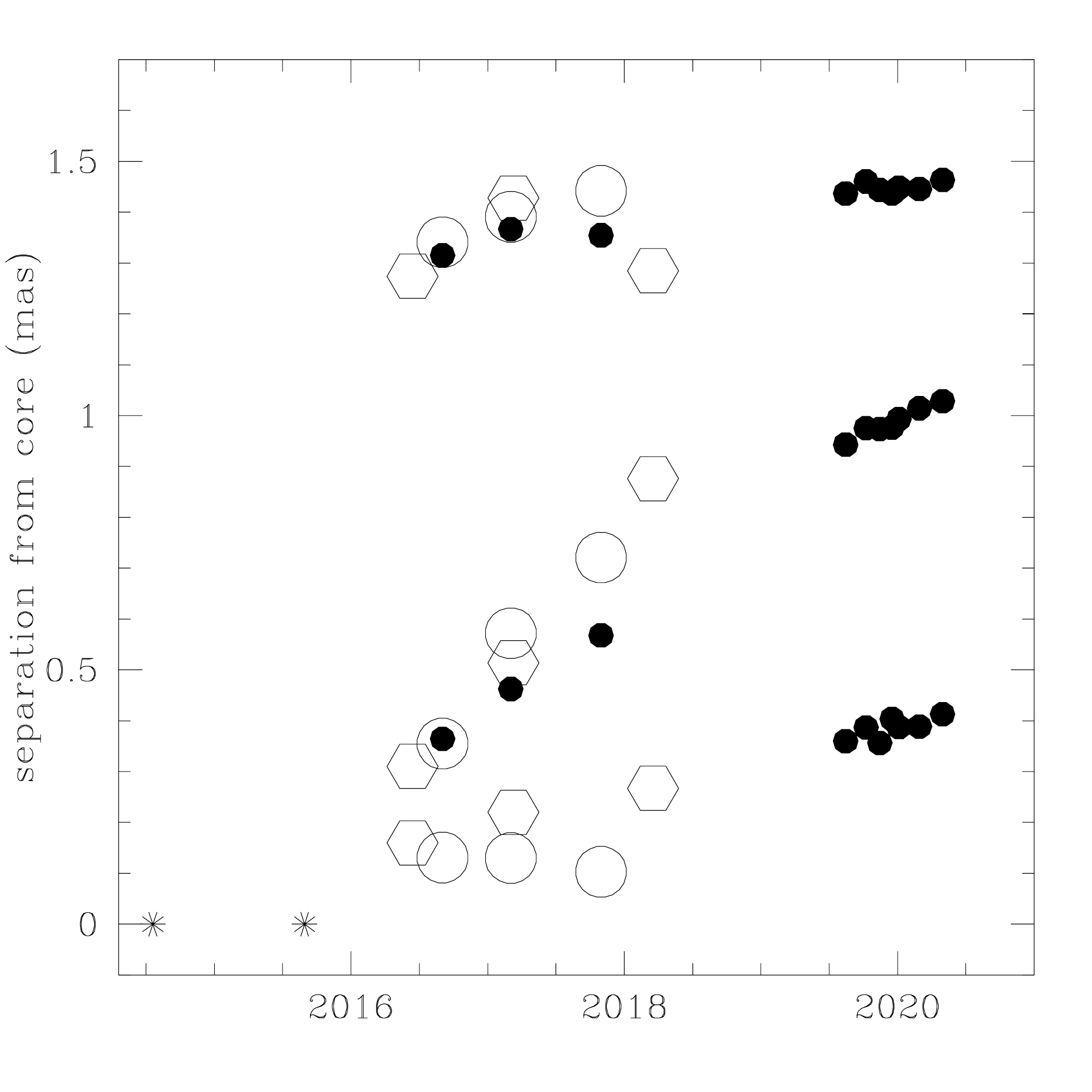}
\caption{The separation between the core and the A (bottom), B (middle), and C (top) components as a function of time. X-axis in years, y-axis in mas; filled circles 15 GHz, open circles 43 GHz, open hexagons 22 GHz. 2015 and 2016 data are from our VLBA and EVN observations, 2019-2020 data from the MOJAVE archive. Asterisks mark the epochs of the two large $\gamma$-ray flares.
}
\label{r5}
\end{figure}

We decided to describe the radio source with a minimum number of Gaussian components. Besides the core, we identified a component A, very close to the core, which is detected only at 22 and 43 GHz in our data, but is well detected in the 15 GHz MOJAVE data taken 4 years later; a component B, detected at all frequencies, definitely moving outwards, and a component C, apparently steady. Both B and C components were required to fit the MOJAVE 2020 data: our C component corresponds to the Ca component in the \citet{2018MNRAS.478..359K} and \citet{2008A&A...483..125R} nomenclature. The uncertainty on our component positions is $\sim$0.1 mas.

We report in Table \ref{vlbi2} for all our components (core, A, B, and C), for each epoch and frequency, the distance from the core, the position angle, the size, and the flux density for VLBA data; these flux values should be considered as rough estimates, because with different resolution and visibility coverage there may be significant blending between compact components and diffuse emission. In particular, we expect a systematically lower flux with respect to the OVRO single dish fluxes reported in our Fig. \ref{curva2}.

\begin{figure}%figure 7
\centering
\includegraphics[height=8cm,angle=0]{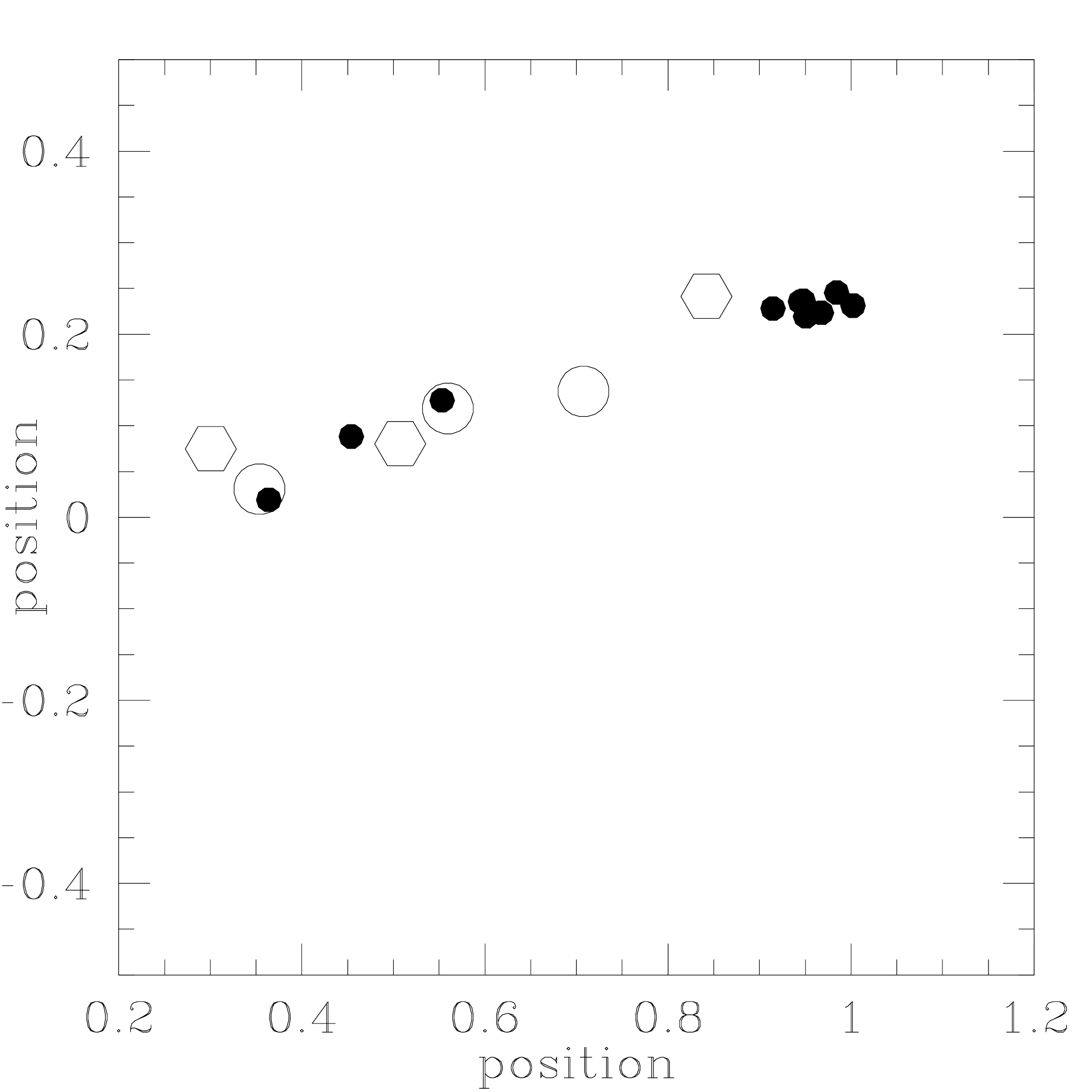}
\caption{The position in the sky plane of the B component at different times with respect to the core at coordinates 0,0. North is up and East to the left. Units are mas for both axis, with 0,0 indicating the core component. Filled circles 15 GHz, open circles 43 GHz, open hexagons 22 GHz.}
\label{r6}
\end{figure}

In Fig. \ref{r5} we plot the evolution of the components' separation from the core with time at the different frequencies: our data are those between 2015 and 2016, while those in the 2019-2020 range come from the MOJAVE archive. The epochs of the two large $\gamma$-ray flares are marked with asterisks located at the core position. In Fig. \ref{r6} we report the positions in the plane of the sky, measured in mas from the core, of component B at different epochs and frequencies: this plot allows to check that the position angle of the motion is consistent with the jet axis, as apparent in Fig. \ref{r4}.
The presence of a quasi stationary, more or less oscillating component at 1.4 mas from the core (our component C), is in agreement with previous studies on this radio source (see e.g. \citealt{2008A&A...483..125R}), while components A and B can be regarded as new. 
Despite the difficulty in relating the components at different epochs with differing UV coverage and observing frequencies, our overall results are consistent with a scenario of a component B with a clear outward motion, with respect to the core component, along the jet direction, with a decreasing flux, and an additional component A separating from the core more slowly and with less reliably estimated positions. 

\begin{figure} %figure 8
\centering
\includegraphics[height=6.4cm,angle=0]{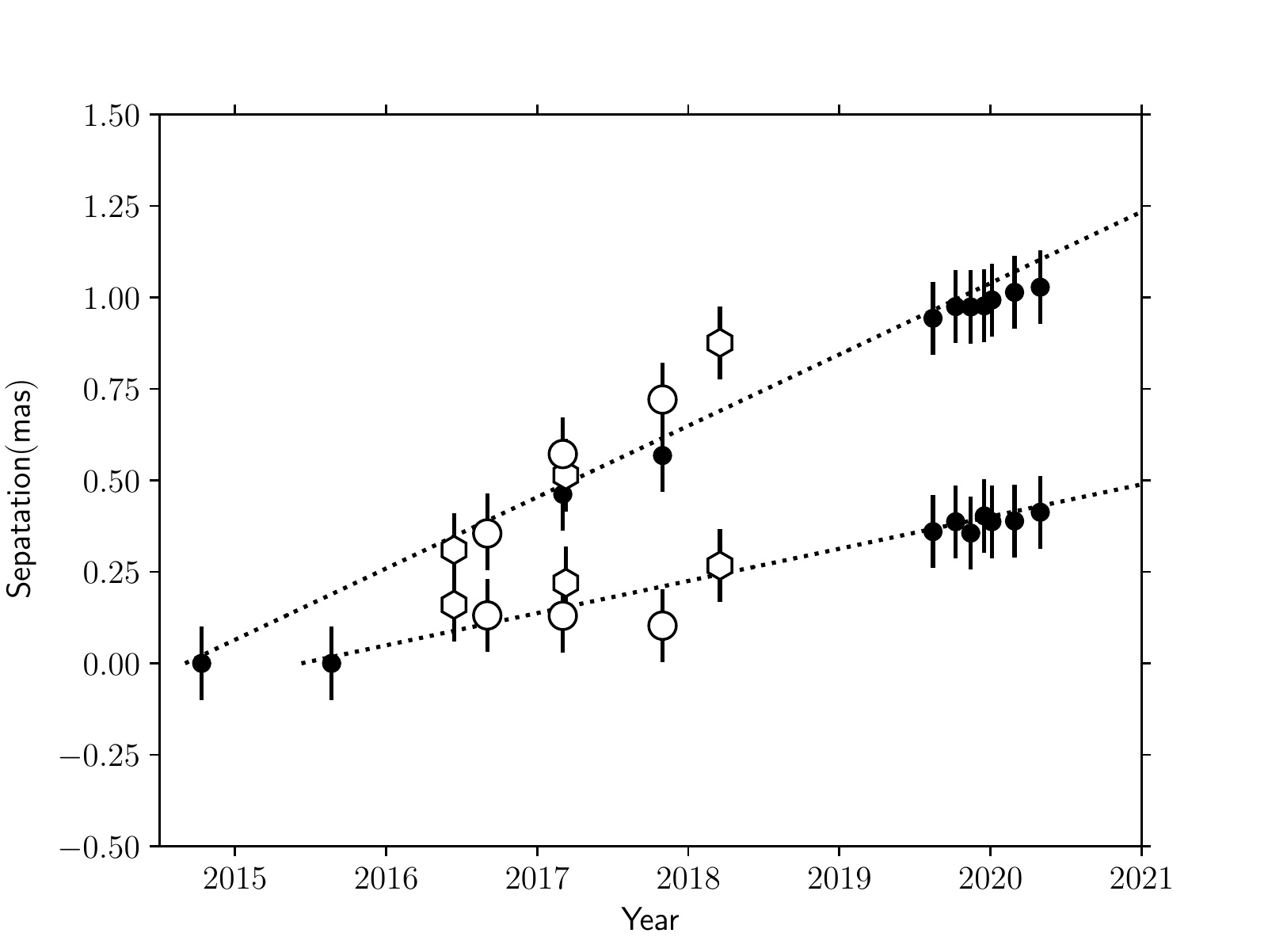}
\caption{The separation between the core and the A (bottom), and B (top) components as a function of time. X-axis in years, y-axis in mas; the regression lines used to evaluate the ejection epochs of the two components are shown. For both components, extrapolation to y=0 are compatible with the times of the two large $\gamma$-ray flares. Filled circles 15 GHz, open circles 43 GHz, open hexagons 22 GHz.
}
\label{r7}
\end{figure}

We derived the proper motion of these components by means of a linear fit (see Fig. \ref{r7}). We found that component A was moving with an apparent velocity v\_app = (0.088$\pm$0.011) mas/yr, corresponding to (4.18$\pm$0.54)c, and the estimated epoch of passage through the VLBI core was 2015.44$\pm$0.44 (i.e. between January and November 2015), consistent with the epoch of the flare on 23 August 2015.

On the other hand, component B was moving with v\_app =(0.195 $\pm$0.014) mas/yr, corresponding to (9.27$\pm$0.67)c, and the estimated epoch of passage through the VLBI core was 2014.67 $\pm$0.28 (i.e. between May and December 2014), consistent with the $\gamma$-ray flare on 19 July 2014.

The OVRO light curve (see Fig. \ref{curva2}) shows a local maximum around the beginning of November 2014 (MJD 56966), that may be related to the ejection of component B. However, the data points around this period are poorly sampled, preventing us from estimating the peak in a more accurate way. The same argument applies to the poorly sampled data points in 2015.

\begin{table}
\caption{VLBI observations epochs. The array (column 1),  the observation project name (column 2),  the date (column 3),  the frequencies used (column 4).}             % title of Table
\label{vlbi1}      % is used to refer this table in the text
\centering                          % used for centering table
\begin{tabular}{c c c c c}        % centered columns (4 columns)
\hline\hline                 % inserts double horizontal lines
array& Project& date& freq.\\
     &        &     & GHz  \\
\hline
EVN& EC057A &14 Jun 2016 &22 \\
VLBA& BC224A& 04 Sep 2016 &15, 43 \\
VLBA& BC224B& 04 Mar 2017 &15, 43 \\
EVN& EC057B& 08 Mar 2017 &22 \\
VLBA& BC224C& 07 Nov 2017 &15, 43 \\
EVN& EC057C &14 Mar 2018 &22 \\
VLBA& MOJAVE& 15 Aug 2019 &15 \\
VLBA& MOJAVE& 08 Oct 2019 &15 \\
VLBA& MOJAVE& 14 Nov 2019 &15 \\
VLBA& MOJAVE& 15 dec 2019 &15 \\
VLBA& MOJAVE& 04 Jan 2020 &15 \\
VLBA& MOJAVE& 28 Feb 2020 &15 \\
 \hline
 \end{tabular}
\end{table}

\begin{table} 
\caption{ Position of radio components with respect to the core: the component nomenclature (column 1), year (column 2) distance from core (column 3), the position angle (column 4),  the size (column 5),  the flux density for VLBA (column 6) and frequency (column 7).}
\label{vlbi2}
\centering                          % used for centering table
\begin{tabular}{c c c c c c c}        % centered columns (4 columns)
\hline\hline                 % inserts double horizontal lines
comp&year& dist. &PA  & size & S &freq.\\
    &    & mas   & deg& mas  &mJy&GHz\\
\hline
core&2016.67& & &0.13&1740&15\\
core&2017.17& & &0.12&1610&15\\
core&2017.83& & &0.11&1720&15\\
core&2019.62& & &0.08&1660&15\\
core&2019.77& & &0.09&1530&15\\
core&2019.87& & &0.10&1470&15\\
core&2019.96& & &0.10&1370&15\\
core&2020.01& & &0.10&1410&15\\
core&2020.16& & &0.10&1430&15\\
core&2020.33& & &0.12&1520&15\\
A&2019.62&0.36&-79&0.27&181&15\\
A&2019.77&0.39&-78&0.29&159&15\\
A&2019.87&0.36&-76&0.41&206&15\\
A&2019.96&0.40&-80&0.28&147&15\\
A&2020.01&0.39&-79&0.32&176&15\\
A&2020.16&0.39&-76&0.39&158&15\\
A&2020.33&0.41&-80&0.39&146&15\\
B&2016.67&0.36&-87&0.32&254&15\\
B&2017.17&0.46&-79&0.41&264&15\\
B&2017.83&0.57&-77&0.51&183&15\\
B&2019.62&0.94&-76&0.31&148&15\\
B&2019.77&0.98&-77&0.33&150&15\\
B&2019.87&0.97&-76&0.33&147&15\\
B&2019.96&0.98&-76&0.33&131&15\\
B&2020.01&0.99&-77&0.35&139&15\\
B&2020.16&1.01&-76&0.32&126&15\\
B&2020.33&1.03&-77&0.31&102&15\\
C&2016.67&1.32&-92&0.52&184&15\\
C&2017.17&1.37&-93&0.62&213&15\\
C&2017.83&1.35&-91&0.77&251&15\\
C&2019.62&1.44&-88&0.31&130&15\\
C&2019.77&1.46&-88&0.36&138&15\\
C&2019.87&1.44&-87&0.36&157&15\\
C&2019.96&1.44&-88&0.39&167&15\\
C&2020.01&1.45&-88&0.38&171&15\\
C&2020.16&1.45&-88&0.42&190&15\\
C&2020.33&1.46&-88&0.45&197&15\\
core&2016.45& & &0.09&-&22\\
core&2017.19& & &0.09&-&22\\
core&2018.21& & &0.08&-&22\\
A&2016.45&0.16&-90&$<$.05&-&22\\
A&2017.19&0.22&-66&0.14&-&22\\
A&2018.21&0.27&-66&0.25&-&22\\
B&2016.45&0.31&-76&0.38&-&22\\
B&2017.19&0.51&-81&0.28&-&22\\
B&2018.21&0.88&-74&0.14&-&22\\
C&2016.45&1.27&-93&0.65&-&22\\
C&2017.19&1.43&-91&0.25&-&22\\
C&2018.21&1.29&-91&0.51&-&22\\
core&2016.67& & &0.06&1110&43\\
core&2017.17& & &0.05&992&43\\
core&2017.83& & &0.06&1110&43\\
A&2016.67&0.13&-46&0.09&144&43\\
A&2017.17&0.13&-69&0.17&115&43\\
A&2017.83&0.10&-126&0.14&80&43\\
B&2016.67&0.36&-85&0.29&64&43\\
B&2017.17&0.57&-78&0.36&61&43\\
B&2017.83&0.72&-79&0.43&83&43\\
C&2016.67&1.34&-92&0.37&83&43\\
C&2017.17&1.39&-93&0.21&55&43\\
C&2017.17&1.44&-92&0.45&90&43\\
\hline
\end{tabular}
\end{table}

\section{Discussion} %section 4
Given the nature of the emission processes producing the spectral energy distribution of a LSP blazar such as S5 1803+784, some correlations are expected between the low and the high frequency components (see e.g. \citealt{2018A&A...616A.172R} and references therein). We briefly discuss here the results of our monitoring.

\subsection{Light curves comparison}
\begin{figure*}%figura 9

\includegraphics[height=6cm,angle=0]{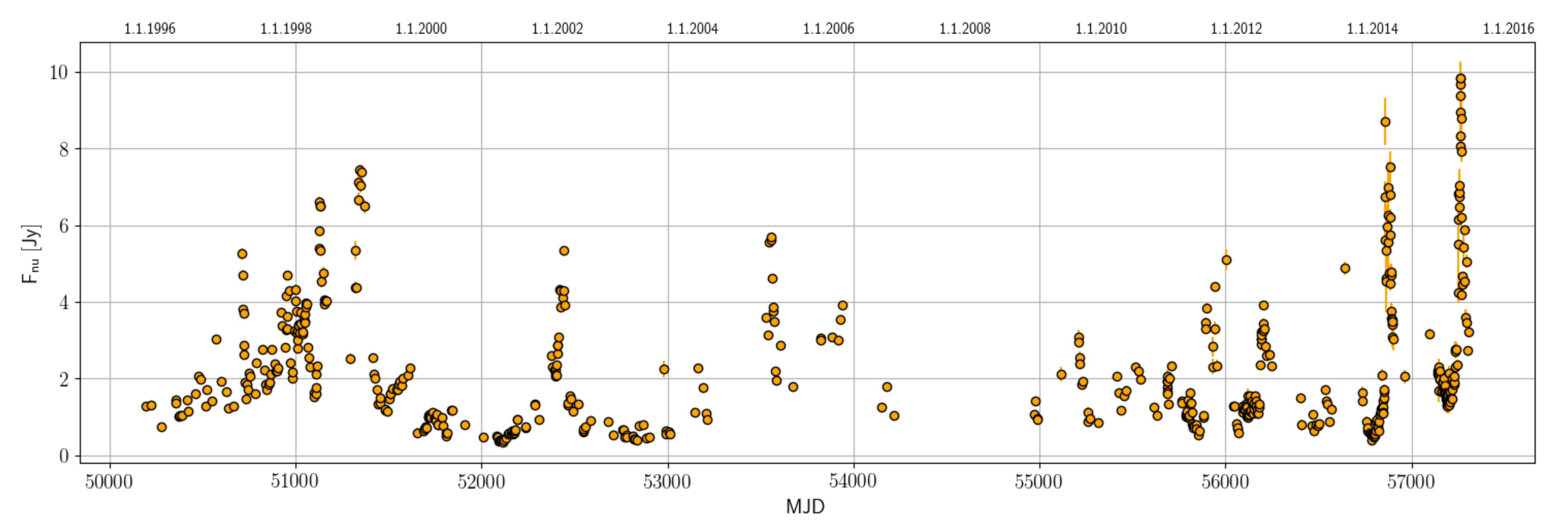}
\caption{The historic R$_C$ light curve of S5 1803+784 from 1996 to 2018.}
\label{1996}
\end{figure*}

The historic optical light curve of S5 1803+784 since 1996 (see Fig. \ref{1996}) shows several flares, but without a well-defined periodicity:  the time interval between  consecutive flares is not constant, nor similar, as can be seen from Fig.\ref{curva2} and Table \ref{opt_flares}. A formal test using the Discrete Fourier Transform (DFT) for unevenly spaced data \citep{1975Ap&SS..36..137D} also gave no dominant frequency.

In the 2008-2018 time interval the source was monitored also by the {\it Fermi}-LAT and had two major flares with a large (factor $\sim$10) variability and a number of minor flares, characterized by a duration of about 60 days. The $\gamma$-ray light curve also does not show a clear periodicity. The main parameters of optical flares  are given in Table \ref{opt_flares}; the $\gamma$-ray ones in Table \ref{Gamma10}. We briefly describe below the main flares. The first flare was followed by us a few days after the $\gamma$-ray flare of 11 January 2010 (MJD 55207) \citep{2010ATel.2386....1D}, so the actual optical peak value is unknown and only the falling slope was measured. A remarkably long high state occurred between MJD 55892 and 56040, peaking at R=14.4 mag; the final part is under-sampled, so the dimming slope is not well constrained. A shorter flare occurred around MJD 56206: the source was bright for about 60 days, but again the dimming phase is not well sampled: apparently no $\gamma$-ray increase was detected. A single bright optical point (R=14.5 mag) was detected by MASTER on 56639, corresponding to a bright $\gamma$-ray level. A very bright flare occurred between MJD 56851 and 56899, apparently  with a small precursor, and a brightening slope of 0.22 mag/day. The source remained in a high activity state for about 45 days, with some oscillations. The actual end was not covered by the monitoring, so the flare was likely much longer. Around MJD 57099 we have just one photometric point, moderately bright (R=15.0 mag), from the Belgium Royal Observatory, corresponding to a high LAT flux. Another very bright flare started at MJD 57248, apparently again with a precursor, with a brightening slope of 0.11 mag/day, followed by an oscillating behaviour. The total duration was about 60 days. Two other minor optical flares occurred on 57348 and 57488, this last one without a $\gamma$-ray counterpart. A long bright state was detected between MJD 57719 and MJD 57747 peaking around R=14.4 mag, with a corresponding high $\gamma$-ray flux. Overall the bright states are typically long-lasting (tens of days) while the rising slopes show a range of values from 0.07 up to 0.22 mag/day. The dimming phases were not well recorded but apparently have flatter slopes.
A fair qualitative correlation between the optical and $\gamma$-ray light curves is apparent from Fig. \ref{curva2}, although the lack of a $\gamma$-ray counterpart for the "moderate" optical flares of October 2012 (MJD $\sim$ 56200) and of April 2016 (MJD $\sim$ 57488) is evident.

\begin{figure}
\centering
\includegraphics[width=1.0\columnwidth]{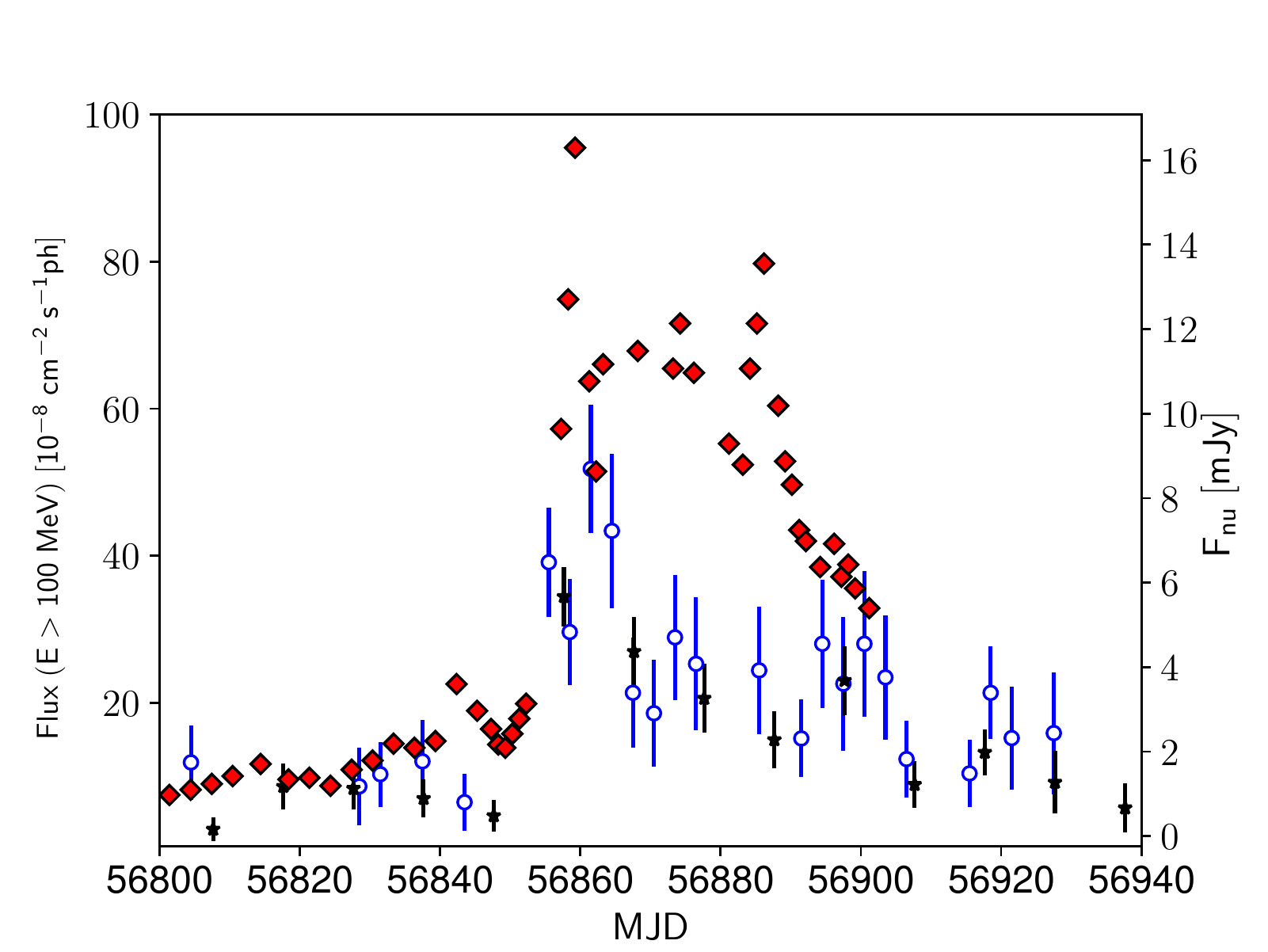}
\includegraphics[width=1.0\columnwidth]{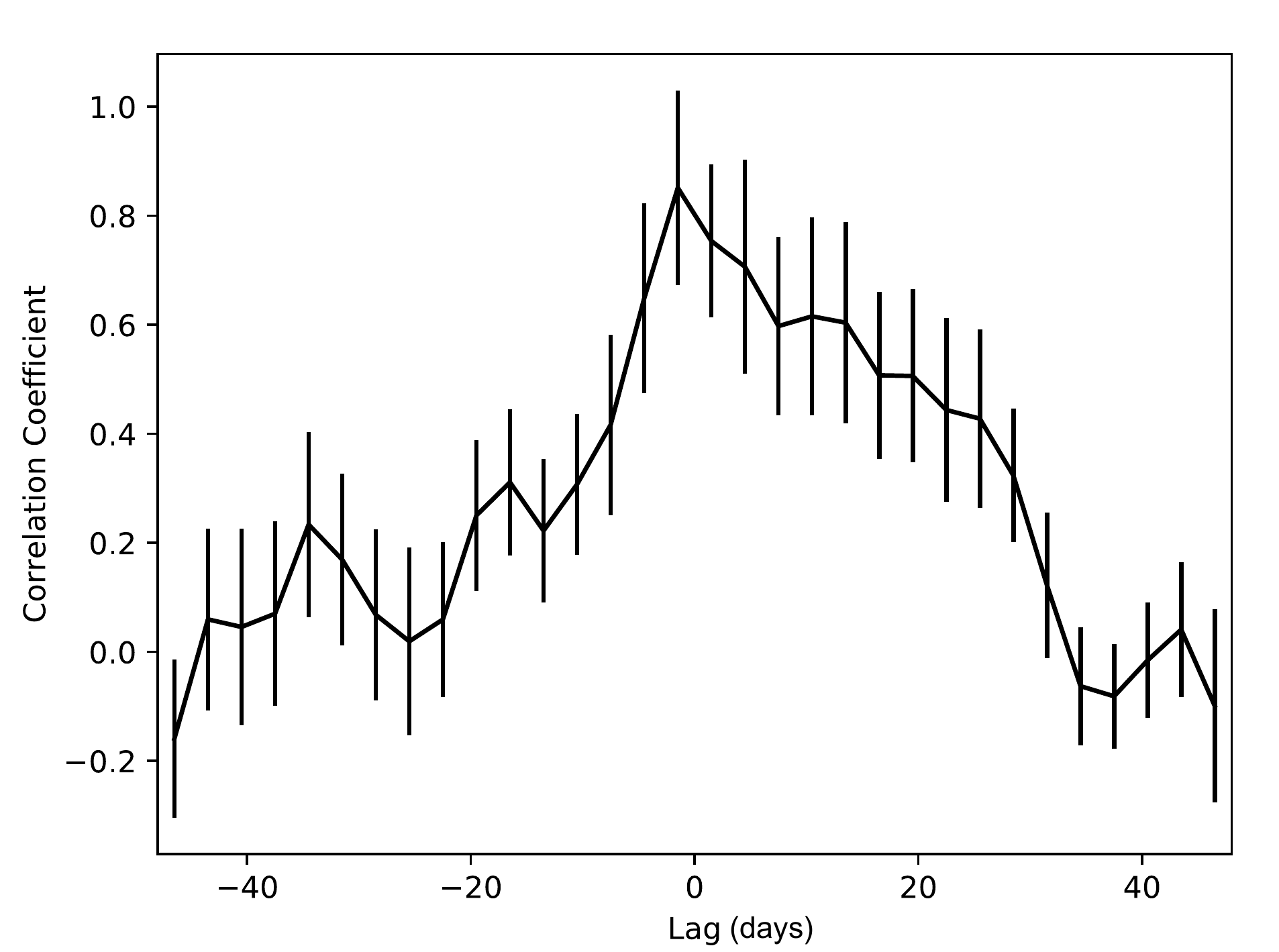}
\hspace{1cm}%\fill}
\caption{Above: the $\gamma$-ray light curve in the 0.1-300 GeV energy range, binned at 1 day and 3 days intervals shown in blue open circle and black star respectively, for the flare around MJD 56857. The upper limit data points are not displayed. The optical light curve is over-plotted for comparison, with the same colour codes of Fig.\ref{curva2}. Below: the Discrete Correlation Function between $\gamma$-ray and optical LC during the first flaring activity: x-axis is in days.}
\label{flare857}
\end{figure}

\begin{figure}
\centering
\includegraphics[width=1.0\columnwidth]{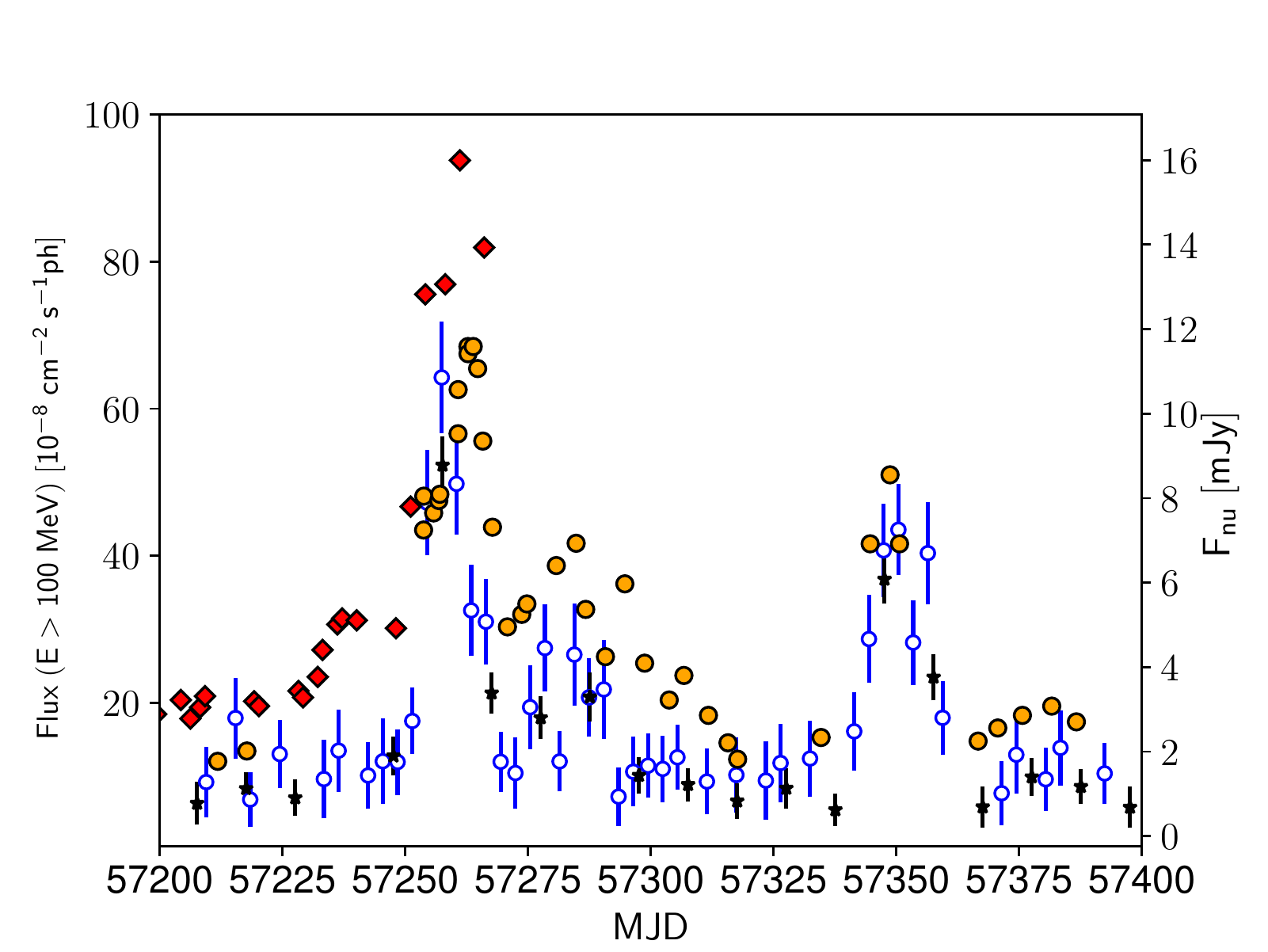}
\includegraphics[width=1.0\columnwidth]{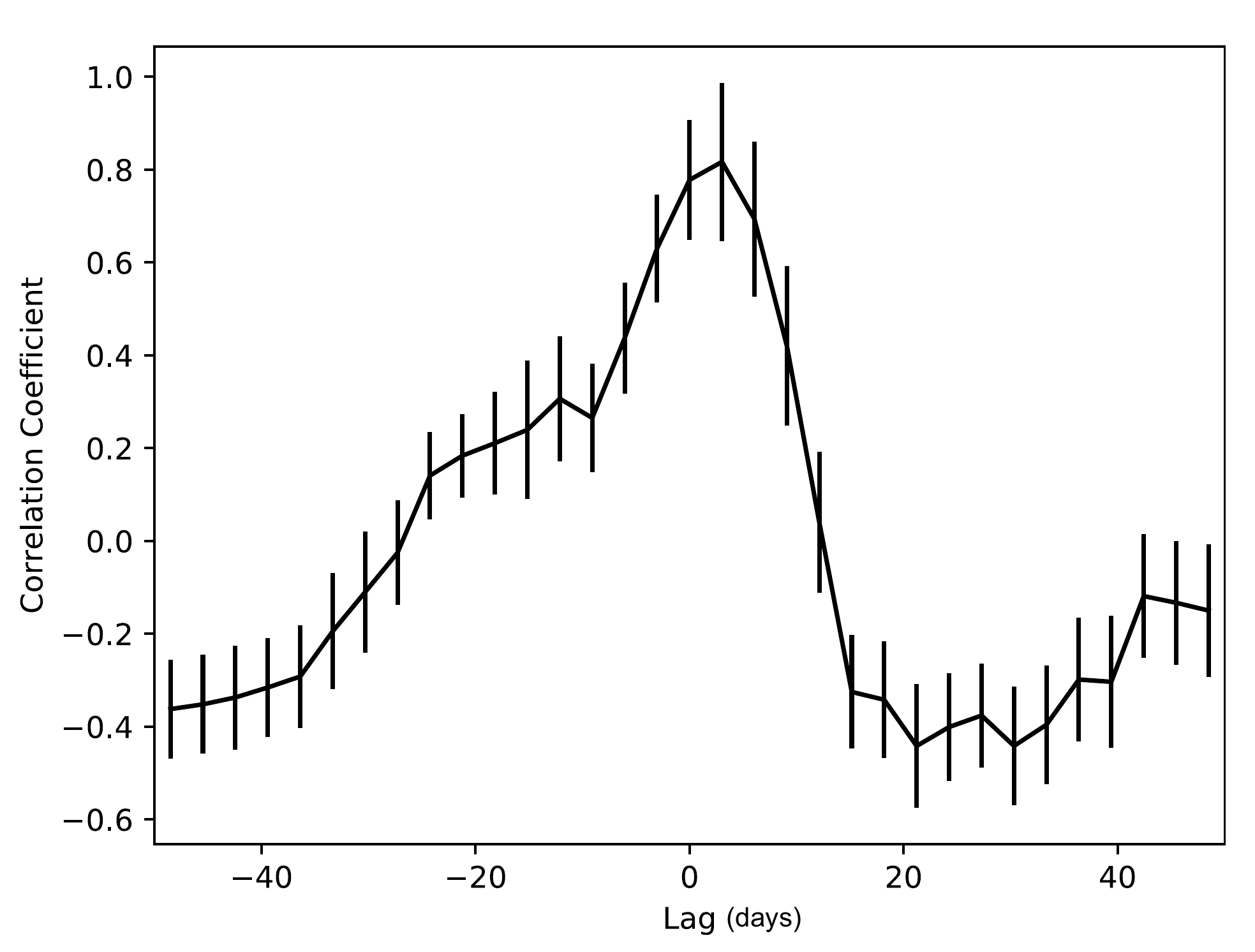}
\hspace{1cm}%\fill}
\caption{Above: the $\gamma$-ray light curve in the 0.1-300 GeV energy range, binned at 1 day and 3 days intervals shown in blue and open circle and black star respectively, for the flare around MJD 57257.  The upper limits data points are not displayed. 
%Grey triangles are upper limits. 
The optical light curve is over-plotted for comparison, with the same color codes of Fig.\ref{curva2}. Below: the Discrete Correlation Function between $\gamma$-ray and optical LC during the second flaring activity: x-axis is in days. Positive values indicate $\gamma$ rays leading.}

\label{flare257}
\end{figure}

For a better comparison, we report zoomed sections of the light curve centered on the last two major flares: Fig.\ref{flare857}  around MJD 56857, and Fig.\ref{flare257} around MJD 57257. The upper panels show the light curves, the lower panels the Discrete Correlation Function (DCF) binned at 3 days intervals.
In the first flare there is a substantial time coincidence of the $\gamma$-ray peak with the optical one. In the second flare, the delay of the optical peak with respect to the $\gamma$-ray one is just 3 days, corresponding to our time bin, so it is not significant.

No clear correlation is present between the 15 GHz total emission, as seen from the OVRO radio telescope, and the optical (and $\gamma$-ray) emission. This is in agreement with the previous finding by \citet{2002AJ....124...53N} based on the Medicina radio telescope at 8.4 GHz in the years 1996-2002.

Regarding the X-ray and optical emissions (see Table \ref{xray-flares}), the overall X-ray flux variation in the {\it Swift} pointings was a factor $\sim$2,  while the corresponding optical luminosities changed by a factor $\sim$6. Overall, no clear correlation emerges from the data, suggesting that the X-ray emission is not directly linked to the optical photons. 
This is expected in a SSC scenario for a LSP BL Lac object, where the X-ray emission is considered to be produced by inverse Compton of relativistic electrons over photons of the radio/submm range:
the available X-ray data are however too sparse to look for a correlation between radio and X-ray emission.

\begin{figure}
\centering
\includegraphics[height=6cm,angle=0]{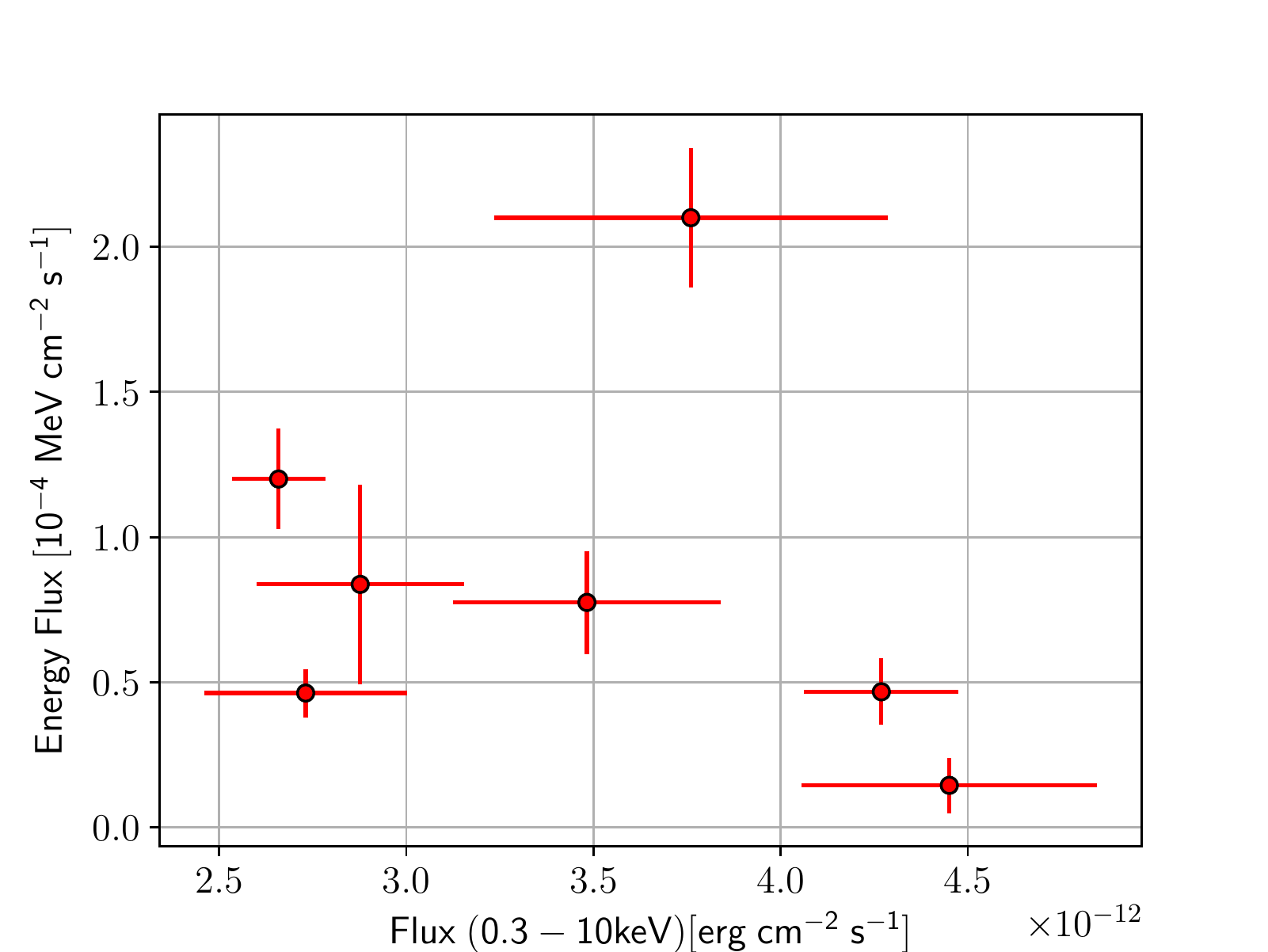}
\caption{The X-ray {\it vs} $\gamma$-ray flux for S5 1803+784 from nearly simultaneous Swift-XRT and {\it Fermi}-LAT data.}
\label{xrtlat}
\end{figure}

We also searched for a correlation between X-ray and $\gamma$-ray fluxes for the available XRT pointings. To this purpose we re-binned the LAT data on 10-day windows centered on the XRT epochs, and summed the XRT data taken at a few days distance: the results are shown in Fig.\ref{xrtlat}.
The range of flux variation in the $\gamma$-ray band was about a factor of 10, against a factor of only $\sim$2 for X-rays. In the SSC scenario the X-ray emission is the low-energy tail of the Inverse Compton energy distribution, which is peaked at $\gamma$-ray frequencies. One may therefore expect that the X-ray flux is larger for larger $\gamma$-ray flux. Unfortunately, we have simultaneous XRT data only during the 2015 $\gamma$-ray flare, while no data are available for the 2014 one. All the other observations were performed with the source in a nearly quiescent state. Our data in Fig.\ref{xrtlat} suggest that, in a quiescent state, there is, if any, an anti-correlation between X-ray and $\gamma$-ray fluxes, i.e. larger X-ray flux for lower $\gamma$-ray flux.

%Also the X-ray photon spectral index does not show a definite trend with the X-ray flux see Table \ref{xray-flares}.

%the values reported near each point in Fig.\ref{xrtlat}).

For S5 1803+784 the position of the $\gamma$-ray peak is around 10$^{22}$ Hz, somewhat below the energy band where the LAT instrument is most sensitive, and therefore not well measured. The LAT data trace the falling branch of the emission, which is expected to be much more variable than the X-ray branch. Therefore the detection of a correlation might require a higher accuracy in the flux measurements for both energy ranges.

\subsection{Spectral slopes}
The optical and UV spectral slopes showed very small variations during our monitoring. The near constancy of the optical and UV color indexes allowed us to more densely build our light curve, as discussed in Sections 2.1 and 2.4. 

The $\gamma$-ray spectrum of the source can be well fitted by a simple power law $N_\nu = A \nu^{-a}$, where N is the number of photons for a given frequency: the photon index $a$ has an average value 2.26$\pm$0.02, which is the transition value between FSRQs and  LSP BL Lacs detected by {\it Fermi}-LAT \citep{2020ApJ...892..105A}. 

We looked for a possible correlation of the photon index with the source luminosity. Indeed we found a
monotonic increase of the spectral index with flux, considering only observations with TS$\ge$25, while this trend disappeared considering all the observations including also the less significance data points.

%We report in Fig. \ref{pi-flux} the photon index {\it vs} the log of the flux, considering only data points with TS$\ge$25. The points with lower statistics (TS$\le$25, not plotted) are mainly located in the upper-left part of the plot and have substantially larger error bars, as expected given their fainter fluxes. This plot suggests that the index monotonically increases with increasing flux up to 1.0$\times$10$^{-7}$ ph~cm$^{-2}$ s$^{-1}$, and hence levels at $\sim$2.2, a "softer when brighter" behaviour. 

 We explored therefore the possibility of a systematic effect in the computation of the spectral index at low flux levels, as is the case of our 10-day time bins. We performed extensive simulations giving as input to the {\it Fermi}-LAT software synthetic spectra with power law indices of 1.5, 2.0, and 2.5, and total fluxes varying as in our source, with the same background as in our real data. Indeed we found that below a flux level corresponding to about 1.0 x 10$^{-7}$ ph cm$^{-2}$ s$^{-1}$, 
the simulated spectra pointed out a systematic underestimation of power law index value.  Therefore the apparent trend found in our observations of S5 1803+784 is  likely  spurious.
%very similar to the case of our source.
%For soft sources with an average spectral index value above 2.0, the computation of ML in a short (10 days) time integration bin does not allow a reliable evaluation of spectral index for low fluxes (1.0 x 10$^{-7}$ ph cm$^{-2}$ s$^{-1}$). 

Finally we built the average spectrum of the quiescent state, summing all the observations with flux below 2.3 10$^{-7}$ ph cm$^{-2}$ s$^{-1}$ (our flaring level defined in Section 2.3), and the average spectrum of the active state, summing all the observations above that level. Both spectra were well described by a power law with the same spectral index (2.22 $\pm$0.01 for the low state and 2.21 $\pm$0.02 for the high state).

\subsection{Optical spectrum}
Besides being useful to measure the redshift, and then the distance of the source, the emission lines in a blazar can help us to understand where they are formed: close to the central engine, before the jet is fully developed, or beyond the jet, illuminated by its radiation. In the first case, if they are excited by the photons coming from the accretion disk, their luminosity should remain unchanged and their E.W. would drop significantly during a flare, which increases only the continuum emission. In the second case, the emission lines are excited by the jet and their luminosity should follow the jet intensity, so their E.W. should remain somewhat constant during a flare.

Our spectrum was taken near a flare peak, with the continuum emission a factor $\sim$9 larger than in 1996 and 2018, when the MgII emission line was evident. If the flux of the emission line were constant, it would be completely undetectable in our spectrum, which is what we actually found. We argue therefore that the emission line region is not excited by the radiation of the jet and is likely close to the central engine.

\section{Conclusions} \label{DA}
No definite periodicity was detected in our optical and $\gamma$-ray light curves of S5 1803+784.
In recent studies \citet{2018MNRAS.478..359K} a period of about 8 years for the precession of the relativistic jet has been claimed, based on the analysis of VLBI images. From the data that we collected, there is no clear optical or $\gamma$-ray counterpart of this phenomenon. 

Regarding the spectral variability, the optical (V,R,I) spectral index was rather constant during our monitoring, regardless of the source optical luminosity, and very similar to that shown by the source in the previous twelve years \citep{2002AJ....124...53N}.  
The X-ray spectral index showed no evidence of hardening at higher fluxes. Also the $\gamma$-ray photon spectral index showed no significant variations correlated with the luminosity, in contrast to  "softer when brighter" behaviour observed in some cases (e.g PKS 2155-304, \citealt{2010ATel.2947....1F} and PKS 0219-164, \citealt{2017ApJ...847....7B}), or a "harder when brighter" behaviour found by \citealt{2013A&A...555A.138F} in the case of PKS 1510-089.
The absence of spectral evolution as a function of the luminosity could be related to the location of the $\gamma$-ray emission region in the flaring state beyond the emission line region. However, being a BL Lacertae object, S5 1803+78 is characterized by extremely weak emission lines, with a small size of the emission line region which does not induce a strong absorption of higher energy $\gamma$-ray photons and a steepening of the spectral index in flaring phase (\citealt{2018MNRAS.477.4749C}).

An optical spectrum taken during the maximum of the 2015 flare did not show the emission line of MgII 2790\AA, present when the source was in a much fainter state: this is expected if the emission line is generated in a volume closer to the core than to the jet emitting the synchrotron radiation, and therefore is not excited by the jet radiation.

We found no correlation between X-ray and optical fluxes: this is expected if the X-ray flux comes from the low-energy tail of an Inverse-Compton emission process, while the optical emission comes from the high-frequency tail of the synchrotron radiation.
On the contrary, the $\gamma$-ray emission showed a fair correlation with the optical one, except in the case of some minor optical flares when no $\gamma$-ray enhancements were detected. In the two major flares, which have a better time sampling, the $\gamma$-ray flux rises simultaneously with the optical one but falls a bit earlier, producing a narrower peak.
The optical/$\gamma$-ray flux ratio is expected to be rather constant in a SSC model: actually, the flux ratios listed in column 5 of Table \ref{Gamma10} show a mildly increasing trend with time, from less than 1 in 2009 to about 3 in 2015, with correlation coefficient 0.84.
The morphology of the source, traced at mas resolution by the VLBI radio observations covering a span of several years after the large $\gamma$-ray-optical flares, shows a clearly detected component moving outwards after the first flare, slowly decreasing in flux. Another similar component appears to have been generated by the second flare, although with a smaller apparent velocity in the plane of the sky. This finding strongly supports a causal connection between the mechanisms producing the high-energy and the radio-band radiation in blazars.

\section*{Acknowledgements}
First of all we want to thank S. Garrappa, S. Buson and D. Thompson for their  insightful  comments,  which improved  the  manuscript  in  its  final  form. We thank the {\it Neil Gehrels Swift Observatory} Time Allocation Committee for the ToO observations of the source during the August 2015 flare.\\ The European VLBI Network is a joint facility of independent European, African, Asian, and North American radio astronomy institutes. Scientific results from data presented in this publication are derived from the following EVN project code: EC057 and from VLBA project code: BC224. The research leading to these results has received funding from the European Commission Horizon 2020 Research and Innovation Programme under grant agreement No. 730562 (RadioNet).\\
The {\it Fermi}-LAT Collaboration acknowledges generous ongoing support from a number of agencies and institutes that have supported both the development and the operation of the LAT, as well as scientific data analysis. These include the National Aeronautics and Space Administration and the Department of Energy in the United States; the Commissariat a l’Energie Atomique and the Centre National de la Recherche Scientifique/Institut National de Physique Nucl\'eaire et de Physique des Particules in France; the Agenzia Spaziale Italiana and the Istituto Nazionale di Fisica Nucleare in Italy; the Centre National d’\'Etudes Spatiales in France; the Ministry of Education, Culture, Sports, Science and Technology (MEXT), High Energy Accelerator Research Organization (KEK), and Japan Aerospace Exploration Agency (JAXA) in Japan; and the K. A. Wallenberg Foundation, the Swedish Research Council, and the Swedish National Space Board in Sweden.\\
Additional support for science analysis during the operations phase is gratefully
acknowledged from the Istituto Nazionale di Astrofisica in Italy and the Centre
National d'\'Etudes Spatiales in France. This work performed in part under DOE
Contract DE-AC02-76SF00515.
MASTER equipment is supported in part by Lomonosov MSU Development Program.
VL is supported by RFBR BRICS grant 17-52-80133.
VK is supported by RFBR grant 19-29-11011.
RRL supports MASTER-IAC observations. \\
This research has made use of public data from the following sources: the OVRO 40-m monitoring program (Richards, J. L. et al. 2011, ApJS, 194, 29) which is supported in part by NASA grants NNX08AW31G, NNX11A043G, and NNX14AQ89G and NSF grants AST-0808050 and AST-1109911; the SIMBAD database, operated at CDS, Strasbourg, France; 
the NASA/IPAC Extragalactic Database (NED) which is operated by the Jet Propulsion Laboratory (JPL), California Institute of Technology, under contract with the National Aeronautics and Space Administration. This research has made use of data from the MOJAVE database that is maintained by the MOJAVE team\footnote{https://www.physics.purdue.edu/MOJAVE/} \citep{2018ApJS..234...12L}.

%%%%%%%%%%%%%%%%%%%%%%%%%%%%%%%%%%%%%%%%%%%%%%%%%%
\section*{Data Availability}

All the data not already included in the Tables are available from the authors on request. 


%%%%%%%%%%%%%%%%%%%%%%%%%%%%%%%%%%%%%%%%%%%%%%%%%%

%%%%%%%%%%%%%%%%% APPENDICES %%%%%%%%%%%%%%%%%%%%%

%\appendix



%%%%%%%%%%%%%%%%%%%%%%%%%%%%%%%%%%%%%%%%%%%%%%%%%%

% Don't change these lines
\bsp	% typesetting comment
\label{lastpage}
\end{document}